\documentclass{IEEEtran}
\IEEEoverridecommandlockouts

\usepackage{booktabs}
\usepackage{amsmath,amsfonts}
\usepackage{algorithmic}
\usepackage{algorithm}
\usepackage{array}
\usepackage[caption=false,font=scriptsize,labelfont=sf,textfont=sf]{subfig}
\usepackage{textcomp}
\usepackage{stfloats}
\usepackage{verbatim}
\usepackage{graphicx}
\usepackage{multirow}
\usepackage{amssymb}
\usepackage{amsthm}
\usepackage{mathrsfs}
\usepackage{indentfirst}
\usepackage{url}
\usepackage{makecell}
\usepackage{float}

\usepackage[normalem]{ulem}
\useunder{\uline}{\ul}{}
\usepackage{xcolor}
\usepackage[bookmarks=false,colorlinks,citecolor=black, filecolor=black, linkcolor=black,urlcolor=black]{hyperref}
\usepackage{eqparbox}
\usepackage{cleveref}
\usepackage[style=ieee, defernumbers=true]{biblatex}
\usepackage{color,url}
\usepackage{wasysym}
\usepackage{threeparttable}
\newcommand{\tabincell}[2]{\begin{tabular}{@{}#1@{}}#2\end{tabular}}
\newcommand\tda{{\rm TDA}}
\newcommand\en{{\rm en}}
\newcommand\de{{\rm de}}
\newcommand\scd{{\rm SC}}
\newcommand\dec{{\rm DEC}}
\newcommand\mlp{{\rm MLP}}
\makeatletter
\let\blx@rerun@biber\relax
\makeatother

\begin{document}

\title{Topology Data Analysis-based Error Detection for Semantic Image Transmission with Incremental Knowledge-based HARQ}

\author{\IEEEauthorblockN{Fei Ni, Rongpeng Li, Zhifeng Zhao and Honggang Zhang}

\thanks{  
   Fei Ni and Rongpeng Li are with the College of Information Science and Electronic Engineering, Zhejiang University, Hangzhou 310027, China (e-mail: \{nifei; lirongpeng\}@zju.edu.cn).
   
    Zhifeng Zhao and Honggang Zhang are with Zhejiang Lab, Hangzhou, China as well as the College of Information Science and Electronic Engineering, Zhejiang University, Hangzhou 310027, China (e-mail: zhaozf@zhejianglab.com, honggangzhang@zju.edu.cn).
  }
}

\maketitle

\begin{abstract}
  Semantic communication (SemCom) aims to achieve high fidelity information delivery under low communication consumption by only guaranteeing semantic accuracy. Nevertheless, semantic communication still suffers from unexpected channel volatility and thus developing a re-transmission mechanism (e.g., hybrid automatic repeat request [HARQ]) is indispensable. In that regard, instead of discarding previously transmitted information, the incremental knowledge-based HARQ (IK-HARQ) is deemed as a more effective mechanism that could sufficiently utilize the information semantics. However, considering the possible existence of semantic ambiguity in image transmission, a simple bit-level cyclic redundancy check (CRC) might compromise the performance of IK-HARQ. Therefore, it emerges a strong incentive to revolutionize the CRC mechanism, so as to reap the benefits of both SemCom and HARQ. In this paper, built on top of swin transformer-based joint source-channel coding (JSCC) and IK-HARQ, we propose a semantic image transmission framework SC-TDA-HARQ. In particular, different from the conventional CRC, we introduce a topological data analysis (TDA)-based error detection method, which capably digs out the inner topological and geometric information of images, so as to capture semantic information and determine the necessity for re-transmission. Extensive numerical results validate the effectiveness and efficiency of the proposed SC-TDA-HARQ framework, especially under the limited bandwidth condition, and manifest the superiority of TDA-based error detection method in image transmission.
\end{abstract}


\section{Introduction}
\label{sec:introduction}
Recently, semantic communication (SemCom), which focuses on transmitting meaningful information rather than precise bits or symbols, emerges as a feasible approach to achieve high fidelity information delivery under lower communication consumption \cite{lu2023semantics,bourtsoulatze2019deep,gunduz2022beyond, strinati20216g}. To achieve superior transmission performance and outstanding channel environmental adaptability, some infant literature adopts joint source-channel coding (JSCC) for the SemCom system. Benefiting from the booming development of deep learning (DL) models in feature extraction and reconstruction, SemCom systems with DL-based JSCC (Deep JSCC) and its extended frameworks are eligible for the transmission of various kinds of data, such as image \cite{bourtsoulatze2019deep, ding2021snr, yang2023witt, grover2019uncertainty, yang2021deep, kurka2020deepjscc, wu2023transformer, lee2023deep, zhang2022wireless, lee2019deep}, speech \cite{weng2021semantic}, text \cite{farsad2018deep, jiang2022deep, zhou2022adaptive} and video \cite{zhai2005joint, jiang2022wireless}, and outperform the traditional design, especially under limited channel bandwidth and low signal-to-noise ratio (SNR). Considering the promising market prospect of computer vision (CV) applications and the foundational role in video transmission, the image SemCom warrants further exploration particularly. 

\begin{figure}[!ht]
\centering
\includegraphics[width=0.35 \textwidth]{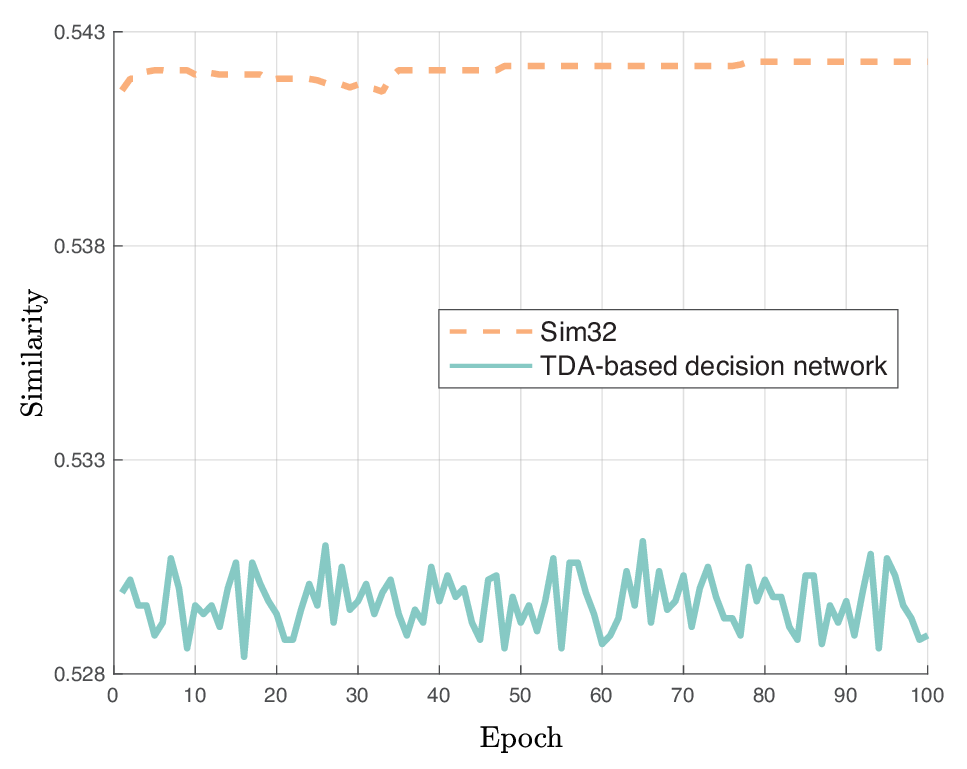}
\caption{The semantic similarity of images detected by Sim32 and TDA-based decision network during the training procedure.}
\label{fig_results-sim32}
\end{figure}

\begin{table}
\caption{The Comparison of error detection results between Sim32 and proposed TDA-based decision network.\label{table_sim32_tda}}
\centering
\begin{tabular}{@{}m{2.3cm}ccc@{}}
\toprule
Method & Similarity & Mean & Variance \\
\midrule
Sim32 & $0.5416 \sim 0.5423$ & $0.5421$ & $3.1151 \times 10^{-8}$\\
\hline
TDA-based decision network & $0.5284 \sim 0.5311$ & $0.5297$ & $4.021 \times 10^{-7}$\\
\bottomrule
\end{tabular}
\end{table}

However, due to the inherent limitations of the traditional DL model's capacity, a notable decline in the performance of Deep JSCC emerges as the image resolution increases.~Meanwhile, as one of the latest variants of transformer architecture, swin transformer shows remarkable performances on many image datasets \cite{liu2021swin}.~Besides, Ref. \cite{yang2023witt}, which utilizes swin transformer as the backbone to realize semantic image transmission, demonstrates considerable performance gain. Motivated by these facts, we adopt swin transformer for the extraction of sophisticated semantic features from the source image, thereby facilitating a more effective method for wireless image transmission.~However, most of the aforementioned frameworks discard useful information in unacknowledged received signals and merely employ a one-shot decoding \cite{lee2023deep}.~Nevertheless, since the lossy information may still embody valuable semantic parts that can assist the receiver in decoding and improving transmission efficiency, the one-shot decoding misses part of the information in rounds of transmissions while reconstructing a high-quality image. 

To combat this deficiency, an incremental knowledge-based hybrid automatic repeat request (IK-HARQ) is introduced for wireless transmission \cite{jiang2022deep, kurka2019successive} and plays an indispensable role in SemCom \cite{zhou2022adaptive}. Specifically, apart from the similar error detection-based acknowledgement/negative acknowledgement (ACK/NAK) feedback to the HARQ in traditional transmission systems, IK-HARQ in SemCom additionally recycles the contaminated information as the valuable incremental knowledge to supplement re-transmission decoding, which further guarantees the accuracy of the received packets \cite{makki2014finite}.
Notably, unlike the bit-level error detection standard (e.g., cyclic redundancy check [CRC] \cite{castagnoli1993optimization}, Hamming Code \cite{hamming1950error}), IK-HARQ in SemCom is expected to discriminate the errors from a semantic perspective, and the corresponding DL-based error detection modules for text transmission in Refs. \cite{jiang2022deep} and \cite{zhou2022adaptive} provide an anspicious start for the future research. However, Ref. \cite{zhou2022adaptive} makes its re-transmission decision solely on channel conditions (i.e., the value of SNR), and lacks a comprehensive understanding of source and environment complexity. Meanwhile, Ref. \cite{jiang2022deep} employs a Sim32 method to calculate semantic similarity at the receiver side and make re-transmission decisions according to a preset threshold. Albeit its effectiveness in semantic text transmission, when this approach is extended to image transmission\footnote{The details to apply Sim32 for computing semantic similarity are given in Appendix.}, the detected similarity tends to be overly stiff (i.e., a variance of $3.1151 \times 10^{-8}$ only) as is illustrated in Fig.~\ref{fig_results-sim32} and Table~\ref{table_sim32_tda}, resulting in re-transmission decisions highly vulnerable to inaccurately configured thresholds. Therefore, boosting the detection robustness necessitates a novel error detection scheme by discovering intrinsic high-level patterns.

\begin{figure}[!t]
\centering
\includegraphics[width=0.495\textwidth]{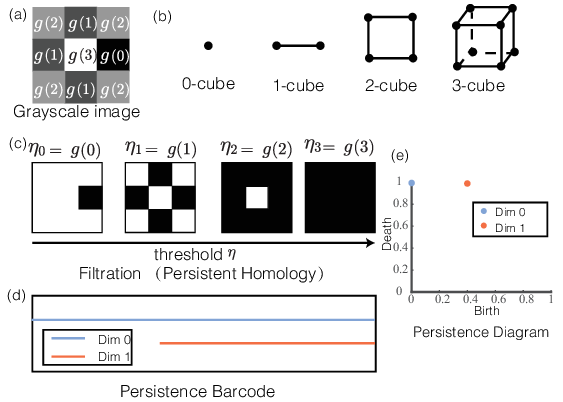}
\caption{An example of a filtered cubical complex's PH. (a) A $3 \times 3$ grayscale image. (b) Different dimension of cubes that constitutes the cubical complex. (c) The filtration process of the image. (d) The corresponding persistence barcode. (e) The corresponding persistence diagram.}
\label{fig_example_PH}
\end{figure}

On the other hand, topological data analysis (TDA), which correlates with the inherent topological patterns of an image \cite{chazal2021introduction,bubenik2015statistical}, is introduced to extract intrinsic features from the source image, demonstrating robustness in error detection as shown in Fig.~\ref{fig_results-sim32} and Table~\ref{table_sim32_tda}. Specifically, TDA leverages topology and geometry to robustly infer qualitative and quantitative information about data structure. In other words, it produces recapitulative summaries or approximations with specific methods like \textit{persistent homology} (PH) \cite{adams2017persistence,balakrishnan2012minimax,bonis2016persistence,cang2017topologynet,chazal2014convergence}. Taking the PH in Fig.~\ref{fig_example_PH} as an example, a $3 \times 3$ images $x$ is constructed by pixels with different grayscale values $g$. Hence, as is shown in Fig.~\ref{fig_example_PH}(b), $x$ can be constructed as the cubical complex $K$ from a union of \textit{multiple} $d \in \{0,1\}$-cubes $\xi$, where $0$-cube and $1$-cube refer to vertices and edges respectively. Given a threshold $\eta$, when its grayscale value $g \leq \eta$, the corresponding pixel spans a $0$-cube. Meanwhile, a $1$-cube emerges when both adjacent pixels have $g \leq \eta$. Thus, different thresholds $\eta$ correspond to different cubical complexes, indicating different topological features. In order to depict how $q$-dimensional topological features persist across different thresholds $\eta_0 < \eta_1 < \eta_2 < \eta_3$, the filtration of $K$ is defined as the nested family of subspaces $K(x,\eta_0) \subseteq K(x,\eta_1) \subseteq K(x,\eta_2) \subseteq K(x,\eta_3)$. Along with the filtration, the topological features evolve, thus affecting the PH. As illustrated in Fig.~\ref{fig_example_PH}(c), $0$-dimensional homology captures the connected components once $\eta$ increases to $g(0)$ while $1$-dimensional homology characterizes a loop that lasts from $\eta = g(1)$ to $\eta = g(3)$. As a result, Fig.~\ref{fig_example_PH}(d) records the corresponding PH as a well-defined set of disjoint half-open intervals known as a \textit{persistence barcode} (PB), where each interval in dimension $q\in\{0,1\}$ symbolizes the lifespan $[b^q, d^q]$ of a topological feature (i.e., $q$-homology class) with the term of the birth time $b^q$ and the death time $d^q$ as the beginning and end of the interval of the $q$-homology. Alternatively, the same information can be represented through a \textit{persistence diagram} (PD), as is shown in Fig.~\ref{fig_example_PH}(e), where intervals are depicted as points $(b^q,d^q)$ in a birth-death coordinate.~The extracted topological and geometric information from images competently describe a complete picture without dimensionality reduction \cite{chazal2021introduction}, and thus potentially amplify the slight variations on core semantic information after noisy transmission.~Inspired by these promising findings \cite{adams2017persistence,balakrishnan2012minimax,bonis2016persistence,cang2017topologynet,chazal2014convergence}, there emerges a strong incentive to develop a TDA-based error detection mechanism and compensate for the deficiency of conventional error detection schemes, so as to fully ensure the gains of IK-HARQ. In this regard, preliminary results in Fig.~\ref{fig_results-sim32} and Table~\ref{table_sim32_tda} validate the more competent capability of the TDA-based scheme.

\begin{table*}[htbp]
\caption{\label{tab_summary}The comparison with highly related literature.}
\centering
\begin{tabular}{c|m{5.3cm}|m{6.5cm}}
\toprule
Related works & Advantages & Limitations \\ 
\hline
Refs. \cite{bourtsoulatze2019deep,ding2021snr,yang2023witt,grover2019uncertainty,yang2021deep} & Superior to traditional systems under limited bandwidth and low SNR. & Contingent on one-shot decoding which discards useful information in unacknowledged signals. \\ \hline
Ref. \cite{jiang2022deep} & \multirow{3}{*}{\tabincell{l}{Using IK-HARQ to combat the deficiency\\ of one-shot decoding.}} & Stiff semantic similarity metrics for image re-transmission decisions. \\ \cline{1-1} \cline{3-3}
Ref. \cite{zhou2022adaptive} & & Making re-transmission decisions based on SNR only and unable to deal with the complexity of delivered content. \\
\midrule
Ours & \multicolumn{2}{l}{\tabincell{l}{Incorporating a TDA-based decision network to check semantic errors in received images \\ and determine re-transmissions.}} \\ \bottomrule
\end{tabular}
\end{table*}

In this paper, on top of IK-HARQ and the state-of-the-art semantic image encoder, swin transformer \cite{liu2021swin}, we focus on developing a TDA-based error detection scheme, and correspondingly design a TDA-based semantic image transmission framework SC-TDA-HARQ. After highlighting key differences with highly related studies \cite{zhou2022adaptive, jiang2022deep, jiang2022wireless} in Table \ref{tab_summary}, SC-TDA-HARQ has the following three-folded merits. 
\begin{itemize}
\item Inspired by the works IK-HARQ \cite{zhou2022adaptive,kurka2019successive}, SC-TDA-HARQ can improve the image transmission efficiency with reduced semantic errors, by effectively leveraging incrementally transmitted knowledge.
\item To further improve the efficiency of semantic image transmission, SC-TDA-HARQ integrates the swin transformer-based semantic coding with an additional TDA coding module to enhance image reconstruction at the receiver side. Therefore, it can exploit the advantages of the semantic architecture and TDA, and outperform competing semantics-based methods in terms of quality of reconstructed images.
\item SC-TDA-HARQ incorporates a TDA-based decision network to determine re-transmissions, by more accurately evaluating the existence of semantic error in estimated images.~In other words, on the condition of semantics consistency, SC-TDA-HARQ can tolerate the reconstruction incorrectness of a few pixels rather than re-transmission, promising to save more communication resources. 
\end{itemize}

The rest of the paper is organized as follows. In Section~\ref{sec:rw}, related work is presented.~The system model and the accompanied framework of SC-TDA-HARQ are introduced in Section~\ref{sec:sm}. In Section~\ref{sec:SC_TDA_HARQ}, we elaborate on the implementation details of SC-TDA-HARQ. The experimental results are demonstrated in Section~\ref{sec:results}, which are followed by our conclusion in Section~\ref{sec:conclusion}.

\section{Related Works}
\label{sec:rw}

\subsection{Deep Joint Source-Channel Coding}
\label{DeppJSCC}
Conventionally, an image transmission system typically uses a separate source-dependent coding scheme (e.g., JPEG \cite{wallace1992jpeg}, JPEG2000 \cite{skodras2001jpeg}, better portable graphics (BPG) \cite{yee2017medical}) to remove the information redundancies.~Subsequently, the transmitted bits undergo a source-independent channel coding scheme (e.g., low-density parity-check (LDPC) \cite{thorpe2003low}, Turbo code \cite{hagenauer1997turbo}) to safeguard from distortions introduced by the noisy communication channel.~Despite the remarkable success and popularity, the independence between the source encoder and the channel encoder may incur restricted transmission performance and a significant reduction in the quality of reconstructed images \cite{bourtsoulatze2019deep}. Thus, JSCC is proposed to overcome the so-called ``cliff effect'' \cite{bourtsoulatze2019deep}, and obtains graceful quality.~The inchoate Deep JSCC architecture for image transmission, whose encoder and decoder are modeled by deep neural networks (DNNs) with excellent generalization ability, directly maps the pixel values to channel inputs and obtains a more reliable reconstruction of source images \cite{bourtsoulatze2019deep}.~Afterwards, many variants of Deep JSCC emerge.~For example, in order to combat channel fluctuations, Ref. \cite{ding2021snr} devises an SNR-adaptive decoder, which uses the pilot signal sent by the transmitter to estimate the SNR and obtains preferable reconstruction quality. Furthermore, Ref.~\cite{yang2023witt} proposes a spatial modulation module to scale the latent representations based on channel state information, thus enhancing the capability to deal with various channel conditions. In addition, inspired by compressed sensing, Ref. \cite{grover2019uncertainty} focuses on reducing the number of measurements corrupted by mild noise and presents uncertainty-aware autoencoders. On the other hand, Ref. \cite{yang2021deep} proposes to combine the orthogonal frequency division multiplexing (OFDM) baseband processing blocks with the Deep JSCC framework, which is superior to traditional schemes and exhibits robustness under the conditions of mismatched channel models as well. 

Notably, all the aforementioned works conduct a \textit{one-shot decoding} method without further iterations. To verify the contribution and the potential of feedback, Ref.~\cite{kurka2020deepjscc} designs practical JSCC schemes that can exploit noisy or noiseless channel feedback, and demonstrates considerable gains as well as excellent adaptability. Ref.~\cite{wu2023transformer} reconfirms the performance of transformer-aided Deep JSCC with feedback. On the contrary, Ref.~\cite{lee2023deep} shows the reconstruction quality improvement can be attained by more iterations at the receiver on the basis of no modifications to the transmitter or requesting re-transmission.~Nevertheless, the performance gains come at the expense of rapidly increased computational expenditure.~In that regard, some feedback and/or re-transmissions sound more appealing.

\subsection{HARQ for Wireless Communication System}
\label{HARQ}
As one of the channel feedback in conventional communication, HARQ can ensure the correctness of the received packets by repetitively re-transmitting until receiving ACK, and is an essential part in a reliable transmission system \cite{jiang2022deep}.~Meanwhile, based on the feedback, adjustable rates can be attained \cite{chande1998joint, lu1999progressive}. Until recently, only a few works notice the latent capability of HARQ in boosting the performance of Deep JSCC. Specifically, in Ref.~\cite{jiang2022deep}, a network called Sim32 is introduced to detect the meaning error in the received sentences.~Concurrently, inspired by Ref.~\cite{kurka2019successive}, Ref. \cite{zhou2022adaptive} pioneers the study of IK-HARQ by regarding previously transmitted lossy information as incremental knowledge to benefit the decoding of re-transmissions. Furthermore, in Ref. \cite{jiang2022wireless}, a semantic video conferencing (SVC) network based on keypoint transmission is established wherein the semantic error detector exploits the fluency of the video to check the received frame. However, faced with the possibility of semantic ambiguity, a convincing error detection scheme remains under-investigated for images.

\begin{figure*}[!ht]
\centering
\includegraphics[width=0.85\textwidth]{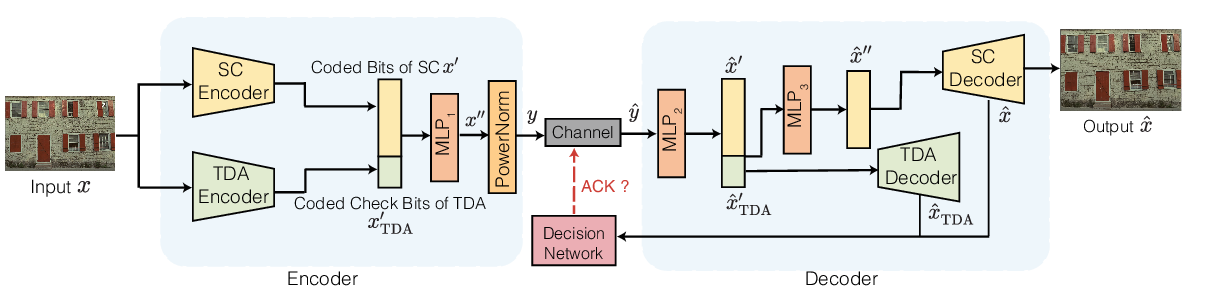}
\caption{The pipeline of the transformer-enabled SemCom system with IK-HARQ and TDA-based error detection.}
\label{fig_pipeline}
\end{figure*}

\begin{figure}[!ht]
\centering
\includegraphics[width=0.495\textwidth]{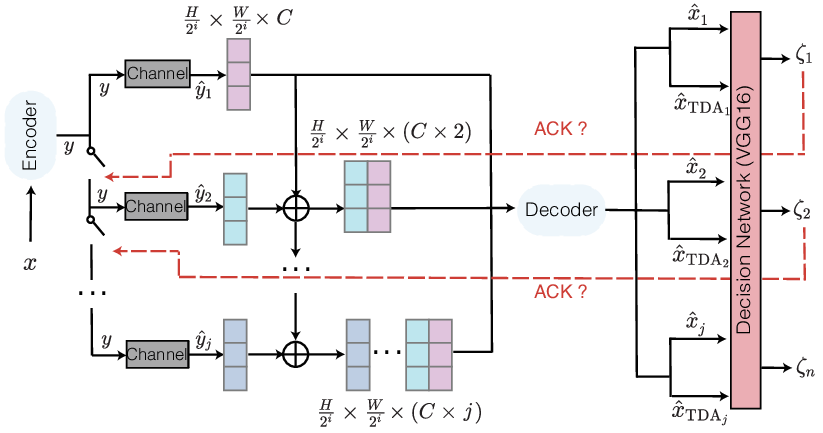}
\caption{The workflow of semantic IK-HARQ at the receiver.}
\label{fig_IK_HARQ}
\end{figure}

\subsection{TDA and its Combination with DL} 
\label{TDA}
Due to its versatility and robustness, TDA has been leveraged to solve various DL problems, after aligning the topological features with the vector input of DL. For instance, Refs. \cite{cohen2005stability} and \cite{mileyko2011probability} introduce Bottleneck distance and Wasserstein distance to statistically compute the similarity of the PDs, thus measuring the topological differences of images.~Alternatively, PB can be transformed into Betti curve, which is defined as the summation of all barcodes belonging to different dimensions and represents the number of topological features \cite{chazal2021introduction}.~Persistence landscape (PL), which focuses on the importance and duration of features during the filtration process \cite{bubenik2015statistical}, belongs to another way to map the PD in a vector space. Besides, in line with the application of kernels in machine learning models, suitable kernel functions are proposed to deal with the topological data as well \cite{reininghaus2015stable, kwitt2015statistical, kusano2016persistence}. Apart from the aforementioned methods that provide the statistic treatment of PD, the topological features can be summarized in terms of persistent entropy (PE) which computes the Shannon entropy based on the probability distribution obtained from the given PD \cite{chintakunta2015entropy}.~On this basis, all previous schemes contribute to significant performance gain in the classification and segmentation tasks of images \cite{adams2017persistence, hofer2017deep, hofer2019learning}. Accordingly, it is promising to introduce TDA to SemCom as a candidate solution to evaluate the credibility of reconstructed images.

\section{System Model}
\label{sec:sm}
\begin{table}[!t]
\caption{Mainly used notations in this paper.\label{table_notation}}
\centering
\renewcommand\arraystretch{1.15}
\begin{tabular}{m{2.1cm}m{5.2cm}}
\toprule
\textbf{Notations} & \textbf{Description}\\
\midrule
$x \in \mathbb{R}^{H \times W \times 3}$ & The source image with height $H$ and width $W$\\
\hline
$\scd_{\en}(\cdot)$, $\scd_{\de}(\cdot)$ & The source-channel encoder and decoder\\
\hline
$\tda_{\en}(\cdot)$, $\tda_{\de}(\cdot)$ & The TDA encoder and decoder \\
\hline
$x^{\prime}$ & The symbol encoded by $\scd_{\en}(\cdot)$\\
\hline
$x_{\tda}^{\prime}$ & The symbol encoded by $\tda_{\en}(\cdot)$ \\
\hline
$x^{\prime \prime}$ & The symbol after compression by MLP $\mlp(\cdot)$\\
\hline
$y$ & Transmitted symbol after power normalization\\
\hline
$\hat{y}$ & The received signal\\
\hline
$\hat{x}_{\tda}$ & The image reconstructed by $\tda_{\de}(\cdot)$\\
\hline
$\hat{x}^{\prime\prime}$ & Recovered symbol by $\mlp(\cdot)$ at the receiver\\
\hline
$\hat{x}$ & The image reconstructed by $\scd_{\de}(\cdot)$ \\
\hline
$\dec(\cdot)$ & The decision network that classifies $\hat{x}$ and $\hat{x}_{\tda}$\\
\hline
$m$ & The source bandwidth \\
\hline
$R$ & The bandwidth compression rate \\
\hline
$\alpha$, $\beta$, $\phi$, $\gamma$ & Trainable parameters for $\scd_{\en}(\cdot)$, $\scd_{\de}(\cdot)$, $\tda_{\de}(\cdot)$ and $\dec(\cdot)$, respectively\\
\hline
$\zeta$ & The ACK/NAK signal computed by $\dec(\cdot)$ \\
\hline
$(u,v)$ & The position of the pixel in images \\
\hline
$g_B(u,v)$ & The binarized function on $(u,v))$ \\
\hline
$\mu_H(\cdot)$, $\mu_R(\cdot)$ & The height and radical filtration on $g_B(\cdot)$\\
\hline
$\psi$, $(u_c,v_c)$ & A candidate direction and center for $\mu_H(\cdot)$ and $\mu_R(\cdot)$ \\
\hline
$b$, $d$ & Birth time and death time of topological features \\
\hline
$q$ & The dimension of the topological features \\
\hline
$t$ & The index of a $q$-dimensional topological feature\\
\hline
$\text{AW}(p,q)$ & The $p$-order Wasserstein amplitude calculated from $(b_t^q,d_t^q)$ \\
\hline
$\text{AB}(q)$ & The Bottleneck distance calculated from $(b_t^q,d_t^q)$ \\
\hline
$\varepsilon(\varsigma,q)$ & The Betti curve of PB at a given threshold $\varsigma$ \\
\hline
$\text{PL}(\lambda,\varrho,q)$ & The persistent landscape for layer $\lambda$ of PD with $\varrho \in \mathbb{N}$\\
\hline
$\text{HK}(\kappa,q,r,s)$ & The heat kernel of PD with respect to $(r,s)\in \mathbb{R}^2$ \\
\hline
$\text{PE}(q)$ & The persistent entropy of PD \\
\hline
$\kappa$ & The standard deviation for Gaussian distribution $\text{HK}(\cdot)$ \\
\bottomrule
\end{tabular}
\end{table}

In this section, we present the system model for semantic image transmission with SC-IK-HARQ and enumerate key components therein, which are further shown in Fig.~\ref{fig_pipeline}. Beforehand, we summarize the mainly used notations in Table.~\ref{table_notation}.
\subsection{Classical Model for SemCom}
\label{sm_sc}
The fundamental SemCom system consists of symmetrical encoder and decoder in the transmitter and receiver. The source-channel (SC) encoder, denoted as ${\scd_{\en}}(\cdot)$, parameterized by $\alpha$ extracts the semantic features of a $3$-color (i.e., RGB) source image $x \in \mathbb{R}^{H \times W \times 3}$ and embeds them into a semantic symbol $x^{\prime}$, where $H$ and $W$ denote the height and width of an RGB image, respectively. Mathematically,
\begin{equation}
x^{\prime} = {\scd_{\en}}(x; \alpha).
\label{eq_SC_en}
\end{equation}

\noindent Then, a multilayer perception (MLP) module, i.e., $\mlp_1(\cdot)$, is used to compress the semantic symbol vector $x^{\prime}$ into transmitted symbol $x^{\prime \prime} \in \mathbb{C}^k$ with the assumption that $k$ is smaller than the source bandwidth $m:=H \times W \times 3$. Subsequently, a power normalization layer \cite{bourtsoulatze2019deep} is then applied to obtain the practically transmitted symbol $y$, that is,
\begin{equation}
y=\sqrt{k} \frac{x^{\prime \prime}}{\sqrt{(x^{\prime \prime})^{\intercal} x^{\prime \prime}}},
\label{eq_power_constraint}
\end{equation}
so as to meet the average power constraint $\mathbb{E}[y^\intercal y] \leq k$. Therefore, as the complex-valued transmission channel enables to transmit two real-valued elements per channel-use \cite{lee2023deep}, the bandwidth compression rate can be written as
\begin{equation}
R = \frac{k}{2m}.
\label{eq_CR}
\end{equation}

Afterwards, $y$ undergoes the communication channel, which can be modeled as an additive white Gaussian noise (AWGN) channel or the Rayleigh fading channel. Correspondingly, the received signal $\hat{y}$ at the decoder can be formulated as 
\begin{equation}
\hat{y} =\begin{cases}
h \cdot y + \theta, & \text{for Rayleigh fading channel}; \\
y+\theta, & \text{for AWGN channel},\\
\end{cases}
\label{eq_channel}
\end{equation}
where $\theta$ is the noise sampled from independent and identically distributed (i.i.d.)~Gaussian distribution $\mathcal{CN}(0,\sigma^2\boldsymbol{I})$ with zero mean and variance $\sigma^2$. In addition, for the Rayleigh fading channel, the channel gain $h$ follows Rayleigh distribution and can be regarded as a constant during a transmission slot. Therefore, the corresponding SNR that evaluates the channel quality can be formulated as $10 \log_{10} \frac{||h \cdot y||^2}{||\theta||^2}$(dB), while AWGN can be viewed as a special case of $h=1$. 

\begin{figure*}[!ht]
\centering
\includegraphics[scale=0.6]{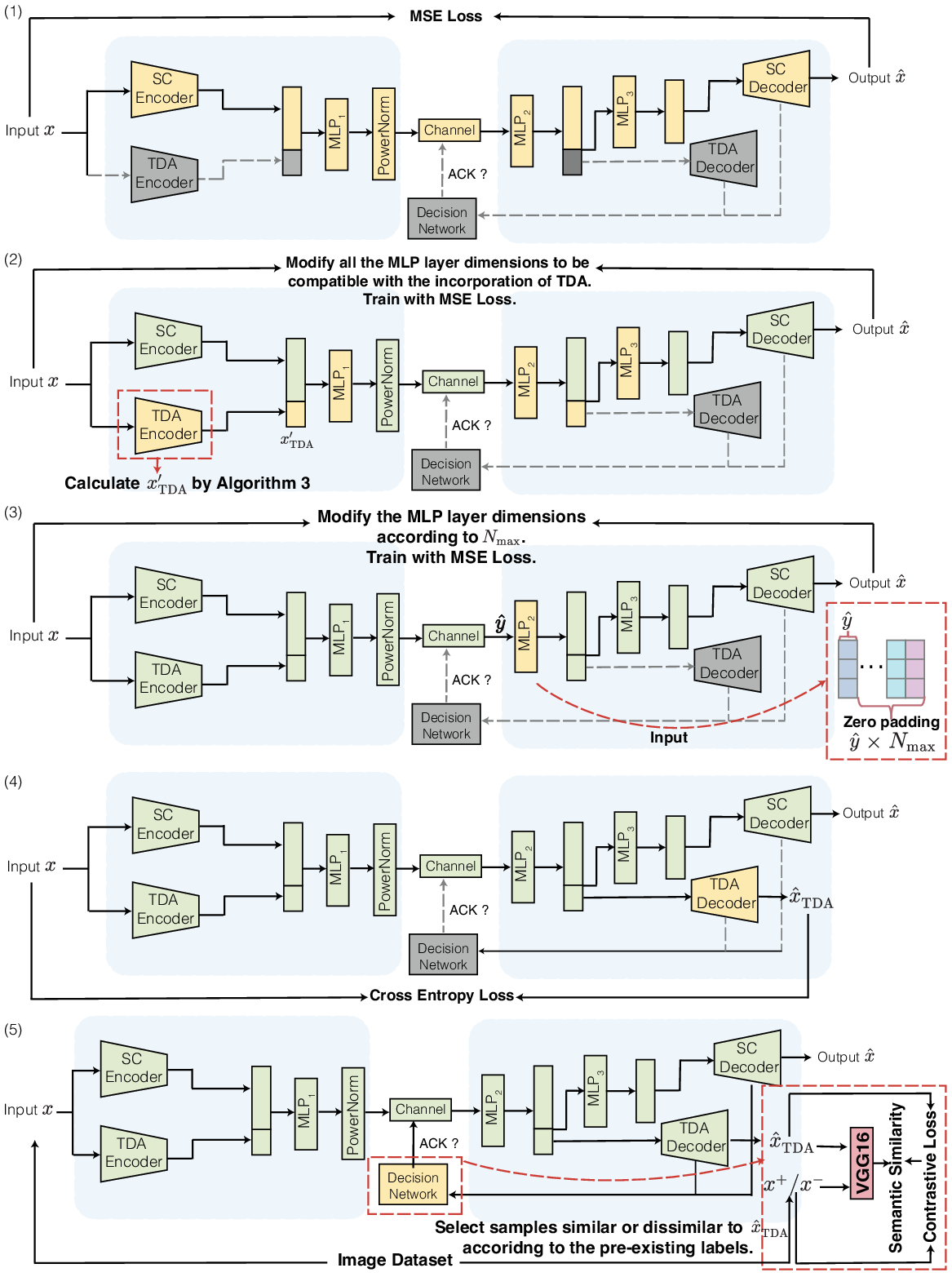}
\caption{The training procedures of SC-TDA-HARQ. Modules colored in yellow are designated to be actively trained. Modules in green indicate that their weights are pre-loaded but not subject to further training during this phase. The modules shown in gray are not included in the training process at this stage.}
\label{fig_training}
\end{figure*}

At the receiver, the channel output $\hat{y}$ undergoes dual MLP modules, so as to obtain a semantic symbol vector $\hat{x}^{\prime\prime}$,
\begin{equation}
    \hat{x}^{\prime\prime} = \mlp_3(\hat{x}^{\prime}) = \mlp_3(\mlp_2(\hat{y})).
\end{equation}
Afterwards, the semantic-channel decoder ${\scd_{\de}}(\cdot)$ with parameters $\beta$ tries to mitigate the physical noise for $\hat{x}^{\prime\prime}$ and recovers the original image in a semantically accurate manner, that is,
\begin{equation}
\hat{x} = {\scd_{\de}}(\hat{x}^{\prime \prime}; \beta).
\label{eq_SC_de}
\end{equation}
\noindent Correspondingly, as is shown in Fig.~\ref{fig_training}(1), a loss function (e.g., mean-squared-error [MSE]) is leveraged to evaluate the distortion between $x$ and $\hat{x}$ and optimize the JSCC parameters $\alpha$ and $\beta$,
\begin{align}
 (\alpha, \beta) & = \arg \min \mathcal{L}_{\text{MSE}} (x,\hat{x}), 
\label{eq_MSE_loss}
\end{align}
where defined by an $L_2$-norm $\Vert \cdot\Vert_2$, $\mathcal{L}_{\text{MSE}} (z,\hat{z})  =  \mathbb{E}  \Vert (z,\hat{z}) \Vert_2$ accompanies with the expectation operator $\mathbb{E}(\cdot)$.

\subsection{TDA-enabled SemCom with IK-HARQ}
\label{sm_harq}

\begin{algorithm}[!h]
\caption{The workflow of IK-HARQ.} 
\label{al:ikharq}
\begin{algorithmic}[1]
\renewcommand{\algorithmicrequire}{\textbf{Initialization:}}
\REQUIRE The transmission iteration indicator $j=1$; the maximum re-transmission times $N_{\max} = 3$; the ACK indicator $\zeta = 1$; the incremental knowledge buffer $y_{\text{buffer}} = \text{NULL}$, well-trained ${\scd_{\en}}(\cdot)$ and $\tda_{\en}(\cdot)$.
\renewcommand{\algorithmicrequire}{\textbf{Input:}}
\REQUIRE The transmitted image $x$.
\renewcommand{\algorithmicrequire}{\textbf{Output:}}
\REQUIRE The reconstructed image $\hat{x}$.
\STATE Compute $y$ based on Eq.~\eqref{eq:fc_1} and Eq.~\eqref{eq_power_constraint} on top of well-trained ${\scd_{\en}}(\cdot)$ and $\tda_{\en}(\cdot)$.
\WHILE{$j \leq N_{\max}$ and $\zeta =0$}
\STATE Update the received vector $\hat{y}_j$ after the $j$-th transmission with Eq.~\eqref{eq_channel}.
\STATE $y_{\text{buffer}} \leftarrow [y_{\text{buffer}},\hat{y}_j] $.
\STATE Update the reconstructed images $\hat{x}^{\prime}$ and $\hat{x}^{\prime}_{\tda}$ with Eq.~\eqref{eq_SC_de} and Eq.~\eqref{eq_TDA_de}.
\STATE Update $\zeta$ according to Eq.~\eqref{eq_ACK}.
\STATE $j = j + 1$.
\ENDWHILE
\end{algorithmic}
\end{algorithm}

Although Deep JSCC has shown effectiveness under low SNR or with the limited bandwidth \cite{bourtsoulatze2019deep}, the veracity of decoded message is still hard to be ensured, especially for the one-shot paradigm.~Consequently, merging SemCom and HARQ is essential to prevent semantic misunderstandings in lossy environments. Furthermore, benefiting from the flexibility in semantics, IK-HARQ allows for efficient information reuse and presents a unified decoder for both transmissions and re-transmissions \cite{zhou2022adaptive}. Compared to classical SemCom, as shown in Fig.~\ref{fig_IK_HARQ} and summarized in Algorithm \ref{al:ikharq}, IK-HARQ introduces a decision network for error detection and revamps the JSCC to accomplish incremental decoding. Therefore, in line with our previous works \cite{zhou2022adaptive}, we implement IK-HARQ as a complement in our image SemCom system.~Notably, we aim to develop a novel TDA-based decision network, instead of traditional methods (e.g., CRC and Sim32), by introducing additional calibration modules and forging the SC-TDA-HARQ. For example, a TDA encoder extracts the corresponding topological features within an image and constructs the inherent relationship between the image and its topological features. Meanwhile, a TDA decoder is necessary to recover the TDA pattern from noisy signals. Next, we enumerate the key components in SC-TDA-HARQ, and highlight essential changes to classical SemCom.

Specifically, at the transmitter, before passing into the MLP module $\mlp_1(\cdot)$, the semantic vector $x^{\prime}$ is first concatenated with the TDA encoding result $x_{\tda}^{\prime}$ by a TDA encoder $\tda_{\en}(\cdot)$, which can be expressed as 
\begin{equation}
x_{\tda}^{\prime} = \tda_{\en}(x).
\label{eq_TDA_en}
\end{equation}
As is depicted in Fig.~\ref{fig_training}(2), all the MLP layer dimensions are modified to be compatible with the incorporation of TDA. In this manner, $x^{\prime \prime} \in \mathbb{C}^k$ can be re-written as
\begin{equation}
    \label{eq:fc_1}
    x^{\prime \prime} = \mlp_1([x^{\prime}, x_{\tda}^{\prime}]).
\end{equation}
Moreover, once receiving an NAK, the re-transmission starts unless the number of re-transmissions reaches $N_{\max}$.

At the receiver, $\hat{y}_{j}$, which denotes the received bits from the $j$-th transmission, is concatenated with previously received ones, followed by another MLP module $\mlp_2(\cdot)$ parameterized by $\rho$ to recover the semantic symbol vector $\hat{x}^{\prime}$ and the topological symbol vector $\hat{x}^{\prime}_{\tda}$. Mathematically,
\begin{align}
(\hat{x}^{\prime}_j, \hat{x}^{\prime}_{\tda_j}) = \mlp_{2}(y_{\text{buffer},j}; \rho),
\label{eq_ik_harq}
\end{align}
where $y_{\text{buffer},j}=[\hat{y}_{1}, \cdots, \hat{y}_{j}]$.~In order to maintain consistent dimension (i.e., $k\times N_{\max}$) of input, $y_{\text{buffer},j}$ in the previous $j<N_{\max} $ transmissions will be padded with zero vectors for decoding, as is illustrated in Fig.~\ref{fig_training}(3).~Therefore, the transmitted knowledge is incrementally leveraged.
Moreover, the MLP module $\mlp_2(\cdot)$ shall be well-trained as
\begin{equation}
    \rho = \arg \min \mathcal{L}_{\text{MSE}} (x, \hat{x}),
\label{eq_fc2}
\end{equation}
so as to capably adapt to different re-transmission times.~Meanwhile, the TDA decoder, denoted as $\tda_{\de}(\cdot)$ with learnable parameters $\phi$, can induce an image $\hat{x}_{\tda}$ according to $\hat{x}^{\prime}_{\tda}$,
\begin{equation}
\hat{x}_{\tda} = \tda_{\de}(\hat{x}^{\prime}_{\tda}; \phi).
\label{eq_TDA_de}
\end{equation}
Furthermore, as is shown in Fig.~\ref{fig_training}(4), the TDA decoder can be trained by
    \begin{align}
        \label{eq_ce1}
        \phi &= \arg \min \mathcal{L}_{\text{CE}} (x, \hat{x}_{\tda}),
    \end{align}   
\noindent where the cross entropy function $\mathcal{L}_{\text{CE}}  (z,\hat{z})  =  \mathbb{E}_z \log \hat{z} $. The pixel values of both images are scaled to effectively transforms each image into a probability distribution, so as to compute the cross entropy loss.

With the recovered $\hat{x}$ and $\hat{x}_{\tda}$, as lately discussed in Section \ref{sec:SC_TDA_HARQ}, a decision network will check the correctness of the decoded bits, and triggers a re-transmission when $\zeta = 0$. Mathematically,
\begin{equation}
\zeta = \dec(\hat{x}, \hat{x}_{\tda}; \gamma)= \begin{cases}
    0, & \text{error detected};\\
    1, & \text{otherwise},
\end{cases}
\label{eq_ACK}
\end{equation}
where $\gamma$ is the trainable parameter in $\dec(\cdot)$. In particular, as demonstrated in Fig.~\ref{fig_training}(5), the decision network, which builds the relationship between an image (e.g. $x$, $\hat{x}$, and $\hat{x}_\tda$) and a decision threshold $\chi_{x}$, can be trained by 
\begin{align}
        \gamma &= \arg \min \mathcal{L}_{\text{Contrastive},\chi_{x}} (x_+/x_-,\hat{x}_{\tda}), \label{eq_ce2}
    \end{align}
where $x^+$ represents images from dataset consisting of $x$ with TDA features matching those of $x_\tda$, and $x^-$ includes images from dataset consisting of $x$ with TDA features different from $x_\tda$.~Moreover, 
the contrastive loss function $\mathcal{L}_{\text{Contrastive},\chi_{x}} (x_+/x_-,\hat{x}_{\tda})$ can be formulated as
\begin{equation}
\begin{aligned}
 &\mathcal{L}_{\text{Contrastive},\chi_{x}} (x_+/x_-,\hat{x}_{\tda})\\
 =&\frac{1}{2}YD^2+\frac{1}{2}(1-Y){\rm max}(0,{\rm margin}-D)^2.
\end{aligned} 
\end{equation}
where $Y$ is a label specified according to $x_+/x_-$ and $x_\tda$. Specifically, when the input pair comprises $x_+$ and $x_\tda$, which indicates that the input belongs to the same category, we set $Y=1$; conversely, for the input $x_-$ and $x_\tda$, we set $Y=0$. In this context, a ${\rm margin}$ is established for these dissimilar pairs to differentiate them more clearly. Additionally, $D$ represents the Euclidean distance between the input pair, serving as a quantitative measure of the similarity or disparity between the two images. On the other hand, $\chi_{x}$ is a decision threshold.
 
Generally, we optimize the parameters $\alpha$, $\beta$, $\phi$ and $\gamma$ for $\scd_\en(\cdot)$, $\scd_\de(\cdot)$, $\tda_\de(\cdot)$ and $\dec(\cdot)$ in a subsequent manner. Consistent with Fig. \ref{fig_training}, a detailed description of this training procedures is presented in Algorithm \ref{al:tda_semcom_ikharq}. On this basis, we will present the details of DNN design and TDA-enabled error detection scheme in Section \ref{sec:SC_TDA_HARQ}.

\begin{algorithm}[!t]
\caption{The training procedures of TDA-enabled SemCom with IK-HARQ.} 
\label{al:tda_semcom_ikharq}
\begin{algorithmic}[1]
\renewcommand{\algorithmicrequire}{\textbf{Initialization:}}
\REQUIRE The maximum re-transmission times $N_{\max} = 3$.
\renewcommand{\algorithmicrequire}{\textbf{Input:}}
\REQUIRE The transmitted image $x$.
\renewcommand{\algorithmicrequire}{\textbf{Output:}}
\REQUIRE The reconstructed image $\hat{x}$.
\STATE Train swin transformer-based ${\scd_{\en}}(\cdot)$ and ${\scd_{\de}}(\cdot)$ by Eq.~\eqref{eq_MSE_loss}.
\STATE 
Add $\tda_{\en}(\cdot)$. Re-train ${\scd_{\en}}(\cdot)$ and ${\scd_{\de}}(\cdot)$ by Eq.~\eqref{eq_MSE_loss}.
\STATE 
Zero padding the already received information $\hat{y}$ according to $N_{\max}$, and train ${\mlp_2}(\cdot)$ by Eq.~\eqref{eq_fc2}.
\STATE Pretrain $\tda_{\de}(\cdot)$ and $\dec(\cdot)$ by Eq.~\eqref{eq_ce1} and Eq.~\eqref{eq_ce2}.
\STATE Optimize the parameters $(\alpha, \beta, \rho, \phi, \gamma)$ by loss function $\mathcal{L}_{\text{MSE}}$.
\end{algorithmic}
\end{algorithm}

\section{The Implementation Details of SC-TDA-HARQ}
\label{sec:SC_TDA_HARQ}
This section elaborates on the implementation details of the proposed image SemCom framework SC-TDA-HARQ, which incorporates IK-HARQ and TDA-based error detection into a classical Deep JSCC paradigm. 

\subsection{Swin transformer-based semantic encoder and decoder}
\label{SC_encoder_decoder}

\begin{figure*}[!ht]
\centering
\includegraphics[width=0.8 \textwidth]{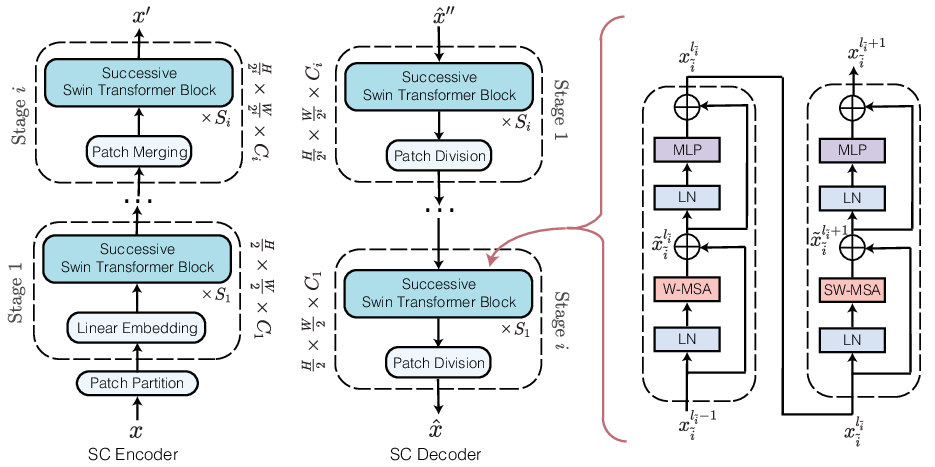}
\caption{The DNN structure of the semantic encoder and decoder for encoding the source image and reconstructing the received signal.}
\label{fig_SC_encoder_decoder}
\end{figure*}

As illustrated in Fig.~\ref{fig_SC_encoder_decoder}, we utilize swin transformer blocks \cite{liu2021swin}, which manifest the excellence in extracting and recovering discriminative semantic representations of images, as the backbone of encoder ${\scd_{\en}}(\cdot)$ and decoder ${\scd_{\de}}(\cdot)$. 

In the encoding process, we stack several stages, each of which embodies a patch merging layer and $S_i$ swin transformer blocks, as the encoder, to project tokens into semantic features. Specifically, as shown in Fig.~\ref{fig_SC_encoder_decoder}, each swin transformer block in a stage fulfills some sequence-to-sequence functionality \cite{liu2021swin}, where a shifted window partitioning approach is adopted by reforming the standard multi-head self-attention (MSA) module with partition windows. 
Consequently, two successive swin transformer blocks $l_{\tilde{i}}$ and $l_{\tilde{i}} + 1$ of stage $\tilde{i} \in \{1,\cdots,i\}$ are computed according to 
\begin{align}
&\tilde{x}_{\tilde{i}}^{l_{\tilde{i}}} = {\rm W} \mbox{-} {\rm MSA}({\rm LN}(x_{\tilde{i}}^{l_{\tilde{i}} - 1})) + x_{\tilde{i}}^{l_{\tilde{i}} - 1}, \nonumber \\
&x_{\tilde{i}}^{l_{\tilde{i}}} = \mlp({\rm LN}(\tilde{x}_{\tilde{i}}^{l_{\tilde{i}}})) + \tilde{x}_{\tilde{i}}^{l_{\tilde{i}}}, \nonumber \\
&\tilde{x}_{\tilde{i}}^{l_{\tilde{i}}+1} = {\rm SW \mbox{-} MSA}({\rm LN}(x_{\tilde{i}}^{l_{\tilde{i}}}) + x_{\tilde{i}}^{l_{\tilde{i}}}, \label{eq_swin_transformer_block}
\\
&x_{\tilde{i}}^{l_{\tilde{i}}+1} = \mlp({\rm LN}(\tilde{x}_{\tilde{i}}^{l_{\tilde{i}}+1}) + \tilde{x}_{\tilde{i}}^{l_{\tilde{i}}+1}, \nonumber
\end{align}
where $\tilde{x}_{\tilde{i}}^{l_{\tilde{i}}}$ and $x_{\tilde{i}}^{l_{\tilde{i}}}$ represent the output feature of the (S)W-MSA module and the MLP module for block $l_{\tilde{i}}$ at stage $\tilde{i}$, respectively.~Notably, $\rm W \mbox{-}MSA(\cdot)$ and $\rm SW \mbox{-}MSA(\cdot)$ are multi-head self attention modules with regular and shifted windowing configurations. The joint application of W-MSA and SW-MSA contributes to learning implicit connections among partitioned windows, and significantly enhances the semantic extraction capability for images.~Besides, $\rm LN(\cdot)$ denotes the layer normalization operation \cite{liu2021swin}. After stage $i$, the source image $x \in \mathbb{R}^{H\times W \times 3}$ is gradually converted into a grid of $\frac{H}{2^i} \times \frac{W}{2^i}$ patches, and subsequently the pixel intensities of each patch are flattened to form a sequence of vectors as a ``token".

Similarly, as illustrated in Fig.~\ref{fig_SC_encoder_decoder}, we adopt a symmetrical DNN structure in the SC decoder, which recovers the image from the noise-polluted representation $\hat{x}^{\prime \prime}$. On this basis, the functionalities in Eqs. \eqref{eq_SC_en} and \eqref{eq_SC_de} can be accomplished.


\subsection{TDA-based Semantic Feature Extraction}
\label{TDA_HARQ}
In this subsection, we talk about the implementation means of the TDA encoder $\tda_\en(\cdot)$, which captures intrinsic topological features of an image $x$, and lays the foundation for a TDA-based error detection scheme.

\begin{figure*}[!ht]
\centering
\includegraphics[width=0.8 \textwidth]{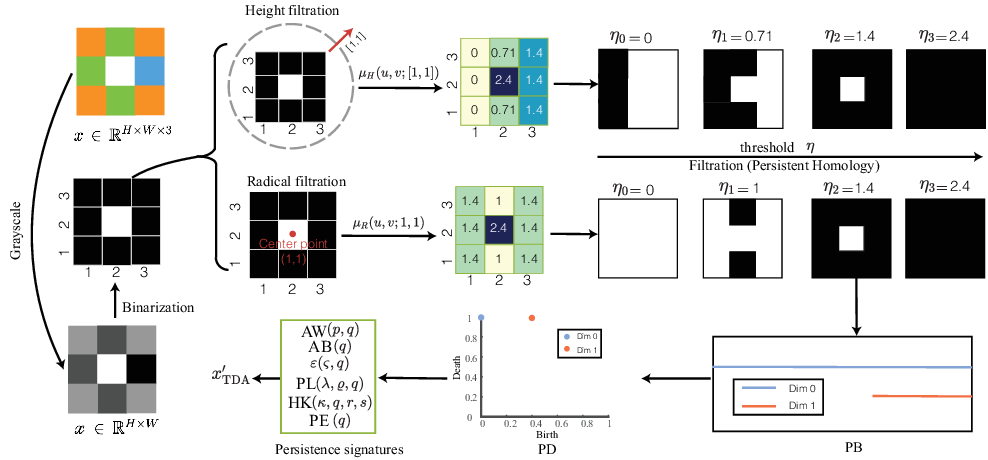}
\caption{An example for the pipeline of the TDA encoder to extract the topological feature. In particular, after grayscale, the source image is binarized according to a preset threshold $\nu$. The height filtration $\mu_{H}(u, v; (1,1))$ and radical filtration $\mu_{B}(u, v; 1,1) $ are further utilized to assign corresponding values to the image. Sequentially, the filtrations are constructed to obtain the PBs and PDs of the topological features. Finally, $x^{\prime}_\tda$ is calculated by vectorizing PDs, in terms of persistence signatures (i.e., the Wasserstein distance $\text{AW}(p, q)$, Bottleneck distance $\text{AB}(q)$, Betti curve $\varepsilon(\varsigma,q)$, persistence landscape $\text{PL}(\lambda,\varrho,q)$, heat kernel $\text{HK}(\kappa,q,r,s)$ and persistent entropy $\text{PE}(q)$).}
\label{fig_TDA}
\end{figure*}

Beforehand, as shown in Fig.~\ref{fig_TDA}, the $3$-color RGB image $x \in \mathbb{R}^{H \times W \times 3}$ is transformed into a grayscale image $\mathscr{x} \in \mathbb{R}^{H \times W}$, with grayscale values $g(u,v)$ at the pixel $(u,v)$ where $1 \leq u \leq H$ and $1 \leq v \leq W$. Then, in order to highlight essential features, $\mathscr{x}$ is binarized according to a preset threshold $\nu$, that is,
\begin{equation}
g_B(u,v) = 
\begin{cases}
1, & \text {if} \ g(u,v)\geq \nu; \\
0, & \text{otherwise}.  
\end{cases}  
\label{eq_binary}
\end{equation}
Recalling the discussions in Section \ref{sec:introduction}, direct grayscale filtration can merely generate few fixed results, which might be unable to comprehensively capture intact topological features.~Therefore, height filtration and radial filtration, which capably incorporate the position information of the homology, are leveraged to make a multiple-perspective analysis of topological features. 
\begin{itemize}
\item \textbf{Height filtration}: Inspired by Morse theory and the PH transform \cite{turner2014persistent}, the height filtration $\mu_{H} : x \rightarrow \mathbb{R}$ of a $2$-dimensional binary image is determined from a chosen unit direction $\psi \in \mathbb{R}^2$ as
\begin{equation}
\mu_{H}(u, v; \psi) = 
\begin{cases}
\langle \frac{\psi}{\Vert \psi\Vert_2}, (u,v) \rangle,  & \text{if} \ g_B(u, v)=1; \\
H_{\infty}, & \text{if} \ g_B(u, v)=0.  
\end{cases}  
\label{eq_height_filtration}
\end{equation}
where $\langle \cdot,\cdot\rangle$ denotes the Euclidean inner product and $H_{\infty}$ is the filtration value of the pixel that is the farthest away from the $\psi$.

\item \textbf{Radial filtration}: Taking the idea from Ref. \cite{kanari2018topological}, the radial filtration $\mu_{R}$ of $\mathscr{x}$ is computed according to the Euclidean distance between a pixel $(u, v)$ and a preset ``center'' pixel $(u_c, v_c)$, that is,
\begin{align}
& \mu_{R}(u, v; u_c, v_c)\label{eq_radial_filtration}
\\
= &
\begin{cases}
\Vert (u_c, v_c) - (u, v)\Vert_2, & \text{if} \ g_B(u, v)=1; \\
R_{\infty}, & \text{if} \ g_B(u, v)=0,  
\end{cases}  \nonumber
\end{align}
where $R_{\infty}$ is the distance in terms of the pixel farthest away from the center.
\end{itemize}
Based on Eq.~\eqref{eq_height_filtration} and Eq.~\eqref{eq_radial_filtration}, by adjusting the direction $\psi$ or the center $(u_c,v_c)$, different filtration values $\mu(u,v)$ (either $\mu_H(u,v)$ and $\mu_R(u,v)$) can be derived from the binary image, thus significantly enhancing the diversity of the captured topological features. Subsequently, a series of cubical complexes is obtained \cite{garin2019topological}. In particular, for a specific filtration $\mu(u,v)$ with the maximum $\mu_{\max}$, the nested family of subspaces satisfies $K(x,\eta_0) \subseteq K(x,\eta_1) \subseteq \dots \subseteq K(x,\eta_n)$, where $0 \leq \eta_0 \leq \cdots \leq \eta_i \leq \cdots \leq \eta_n \leq \mu_{\max}$ and the cubical complex $K(x,\eta_i)$ is formed with a set of cubes $\xi_q$, $q \in \{0, 1, 2\}$. During the filtration, all emerging cubes are sorted as follows: Initially, the $0$-cubes (i.e., the nodes) are randomly sorted. For a $d_1$-cube $\xi_{d_1}$ and a $d_2$-cube $\xi_{d_2}$, the cubes are first sorted by the ascending order of $d_1$ and $d_2$. While for $d_1 = d_2$, the order of cubes is determined by its longest (most specific) $1$-cube (i.e., the edge). As for those equal longest edges (i.e., their longest edges emerge at the same value $\eta$), the cubes are sorted in ascending order of the maximum number of nodes. Subsequently, along with the variations of $\eta \in [0,\mu_{\max}]$, we can achieve a boundary matrix $\tau_{\text{filtration}}$ in which each row and column represents all ordered emerging cubes, as well as an incident matrix $\varpi_{\text{filtration}}$ recording emerging thresholds $\eta$ for all cubes. Furthermore, if the $d_{q-1}$-cube $\xi_{d_{q-1}}$ is the face of the $q$-cube $\xi_q$, the corresponding entry in $\tau_{\text{filtration}}$ refers to $1$, otherwise null \cite{otter2017roadmap}. 

Next, by reducing $\tau_{\text{filtration}}$ into a column-echelon form $\tau_{\text{reduced}}$, the $t$-th topological feature (i.e., the PD $(b^q_t,d^q_t)$) of dimension $q$ can be obtained by simply scanning each column from left to right \cite{otter2017roadmap}. Specifically, for a column contains non-zero entries (i.e., 1), when the row index corresponding to the last $1$ equals the column index, it implies that the cube disappears. Therefore, the corresponding death time $\eta_{\text{death}}$ is the same as the exact birth time for the cube corresponding to the row index and can be found by looking up $\varpi_{\text{filtration}}$. Otherwise, if the values of a column (i.e. the index of the ordered cubes) are all $0$, we resort to lookup the emerged threshold $\eta_{\text{birth}}$ of the cubes, and regard $[\eta_{\text{birth}}, +\infty]$ as the interval of the cube. Correspondingly, the PB for a cube $t$ can be represented as $[b^q_t,d^q_t]=[\eta_{\text{birth}}, \eta_{\text{death}}]$, while the corresponding PD can be denoted as $\vec{l_t^q}=(b^q_t,d^q_t)$. 

\begin{algorithm}[t]
\caption{The TDA-based encoder.}
\label{al:tda_en}
\begin{algorithmic}[1]
\renewcommand{\algorithmicrequire}{\textbf{Initialization:}}
\REQUIRE The number of the directions used in the height filtration $N_{\psi}$; the number of the centers used in the radial filtration $N_c$; the binarized threshold $\nu$.
\renewcommand{\algorithmicrequire}{\textbf{Input:}}
\REQUIRE The transmitted image $x$.
\ENSURE The topological feature $x^{\prime}_{\tda}$.
\STATE Transform the image $x \in \mathbb{R}^{H \times W \times 3}$ into a grayscale image $\mathscr{x} \in \mathbb{R}^{H \times W}$, with grayscale values $g(u,v)$ at a pixel $(u,v)$ $1 \leq u \leq H, 1 \leq v \leq W$. 
\STATE Binarize $\mathscr{x}$ with Eq.~\eqref{eq_binary}. 
\WHILE{$1 \leq s \leq N_\psi$}
\STATE Update pixel value $\mu(u,v)$ with Eq.~\eqref{eq_height_filtration}. 
\STATE Construct the filtration of nested cubical complex $K$ with $K(x,\eta_0) \subseteq K(x,\eta_1) \subseteq \dots \subseteq \subseteq K(x,\eta_i)  \subseteq \dots \subseteq K(x,\eta_n)$, where $0 \leq \eta_i \leq \mu_{\max}(u,v)$. 
\STATE Record all the cubes $\xi$ emerged during the filtration and their emerged threshold $\eta_\xi$. 
\STATE Sort the recorded cubes and construct a corresponding boundary matrix $\tau_{\text{filtration}}$ and an incident matrix $\varpi_{\text{filtration}}$.
\STATE Reduce $\tau_{\text{filtration}}$ as $\tau_{\text{reduce}}$ and compute $(b^q_t, d^q_t)$ according to $\tau_{\text{reduce}}$ and $\varpi_{\rm filtration}$.
\STATE Transform $(b^q_t, d^q_t)$ into the vector input with Eq.~\eqref{eq_distance_w}--Eq.~\eqref{eq_pe}.
\ENDWHILE
\WHILE{$1 \leq s \leq N_c$}
\STATE Update pixel value $\mu(u,v)$ with Eq.~\eqref{eq_radial_filtration}. 
\STATE Repeat steps in Line 5 to Line 9. 
\ENDWHILE
\STATE Update $x^{\prime}_{\tda}$ with Eq.~\eqref{eq_x_tda}.
\end{algorithmic}
\end{algorithm}

Based on the computed PD from height filtration and radial filtration, the PD can be further processed. Correspondingly, consistent with the discussions in Section \ref{TDA}, persistence signatures like the Wasserstein distance, Bottleneck distance, Betti curve, PL, and heat kernel are further adopted. 

\begin{itemize}
\item \textbf{Wasserstein amplitude} \cite{mileyko2011probability}: Based on the Wasserstein distance, the Wasserstein amplitude of order $p$ is defined according to $L_p$ norm of the vector, that is
    \begin{equation}
        \text{AW}(p, q) = \frac{\sqrt{2}}{2} \left( \sum\nolimits_t (d_t^q - b_t^q)^p\right)^{\frac{1}{p}}.
    \label{eq_distance_w}
    \end{equation}
    In this paper, we use $p \in \{1, 2\}$.
\item \textbf{Bottleneck distance} \cite{cohen2005stability}: The Bottleneck amplitude can be considered as a special case of $p \rightarrow \infty$ in \eqref{eq_distance_w}. In other words,
    \begin{equation}
        \text{AB}(q) = \frac{\sqrt{2}}{2} \sup_t (d_t^q - b_t^q).
    \label{eq_distance_b}
    \end{equation}
\item \textbf{Betti curve} \cite{chazal2021introduction}: The Betti curve of a PB is the bar number in $q$ dimension at a given threshold $\varsigma$
    \begin{equation}
            \varepsilon(\varsigma,q) = \sum\nolimits_t \Pi_{[b_t^q,d_t^q]}(\varsigma),
            \label{eq_betti_curve}
    \end{equation}
where $\Pi_{[b_t^q,d_t^q]}(\varsigma) = \begin{cases}
    1, & \varsigma \in [b_t^q,d_t^q];\\
    0, & \text{otherwise}.
\end{cases}$ \cite{pun2018persistent}.
\item \textbf{Persistent landscape} \cite{bubenik2015statistical}: The PL is characterized by a sequence of $\text{PL}(\lambda,\varrho,q) : \mathbb{R} \rightarrow [0, \infty)$ for layers $\lambda=1,2,\dots, \varrho \in \mathbb{N}$, which denotes the $\lambda$-th largest value of $\{f(\varrho,\vec{l_t^q},q)\}_{t=1}^{N_q}$, with $N_q$ indicating the number of bars in dimensions $q$ and
    \begin{equation}
    f(\varrho,\vec{l_t^q},q) = 
    \begin{cases}
    0, & \text{if} \ \varrho \notin (b_t^q,d_t^q); \\
    \varrho-b_t^q, & \text{if} \ \varrho \in (b_t^q, \frac{b_t^q+d_t^q}{2});  \\
    d_t^q-\varrho, & \text{if} \ \varrho \in (\frac{b_t^q+d_t^q}{2},d_t^q).
    \end{cases}  
    \label{eq_pl}
    \end{equation}
Here we consider curves obtained by setting $\lambda \in \{1,2\} $.
\item \textbf{Heat kernel} \cite{reininghaus2015stable}:  A stable multi-scale kernel bridges with machine learning techniques by Gaussian distributions with mean $\iota=(b_t^q, d_t^q)$ and standard deviation $\kappa$ (i.e., $\mathcal{N}(\iota,\kappa)$) \cite{reininghaus2015stable}. Mathematically, denoting $(r,s) \in \mathbb{R}^2$ as any position in the final heat map, the value of $(r,s)$ corresponds to the summation of the output through the Gaussian distribution over every point of the PD. $\text{HK}$ is a real-valued function on $\mathbb{R}^2$ as Eq.~\eqref{eq_hk} on Page \pageref{eq_hk} where $\kappa \in \{10, 15\}$.
\begin{figure*}
    \begin{equation}
        \text{HK}(\kappa,q,r,s) = \frac{1}{8 \pi \kappa} \sum\nolimits_t e^{-\frac{(r-b_t^q)^2+(s-d_t^q)^2}{8 \kappa}}-e^{-\frac{(r-d_t^q)^2+(s-b_t^q)^2}{8 \kappa}},
    \label{eq_hk}
    \end{equation}
    \hrulefill
\end{figure*}

\item \textbf{Persistent entropy} \cite{chintakunta2015entropy}: The PE of a PB takes the Shannon entropy of the persistence (lifetime) of all cycles \cite{atienza2019persistent} 
    \begin{equation}
        \text{PE}(q) = -\sum \nolimits_t \frac{\vec{l_t^q}}{L} \log (\frac{\vec{l_t^q}}{L}),
    \label{eq_pe}
    \end{equation}
    where $L = \sum \nolimits_t \vec{l_t^q}$ is the sum of all the persistences.
\end{itemize}

\begin{figure*}
\begin{equation}
    x^{\prime}_{\tda_s}(f) = [\varepsilon(\varsigma,q), \text{PL}(\lambda,\varrho,q), \text{HK}(\kappa,q,r,s), \text{AW}(p,q), \text{AB}(q), \text{PE}(q)]_f,
    \label{eq_x_tda_spe}
\end{equation}
\hrulefill
\begin{equation}
x^{\prime}_{\tda} = [x^{\prime}_{\tda_s}(\psi_1), \cdots, x^{\prime}_{\tda_s}(N_{\psi}), x^{\prime}_{\tda_s}((u_c, v_c)_1),\cdots x^{\prime}_{\tda_s}((u_c, v_c)_{N_c})]. \label{eq_x_tda}
\end{equation}
\hrulefill
\end{figure*}
For a specific case of filtration $\psi$ or $(u_c,v_c)$, we concatenate all abovementioned metrics as the topological persistence signatures as Eq.~\eqref{eq_x_tda_spe} on Page~\pageref{eq_x_tda_spe}, where the operator $[\cdot]_f$ implies all components are computed under the filtration parameterized by $f$ (e.g., $\psi$ and $(u_c,v_c)$). 
Furthermore, after the process of the TDA encoder, $x^{\prime}_{\tda}$ can be achieved as Eq.~\eqref{eq_x_tda} on Page \pageref{eq_x_tda} by further concatenating Eq.~\eqref{eq_x_tda_spe} from $N_{\psi}$ height filtration and $N_c$ radial filtration. Meanwhile, the procedures for TDA encoding have been summarized in Algorithm~\ref{al:tda_en}.

\subsection{TDA-based Decision Network for Error Detection}
The stability theorem \cite{cohen2005stability} proves that two similar objects possess close PDs. Consequently, it is eligible to use PD or the induced persistence signatures to measure the topological closeness between two images. Therefore, intuitively, the topological features of the reconstructed image $\hat{x}$ at the receiver can be computed for the ACK/NAK decision making, by comparing with the transmitted TDA features, named as TDA-Computation decision-making approach. Alternatively, topological features can be regarded as a special type of hidden features and source images can be recovered accordingly by utilizing various methods like generative adversarial network (GAN) \cite{goodfellow2014generative}, convolutional neural network (CNN). Nevertheless, both means can be considerably time-consuming, and it is worthy of investigation of a computation-effective means for making error detection decisions. 

\begin{table}[!tb]
\caption{\label{tab_RT}Comparison of PSNR and running time between TDA-Recovey and TDA-Computation decision-making approaches.}
\centering
\begin{tabular}{ccc}
\toprule 
Method & PSNR & Running time \\
\midrule
\specialrule{0em}{1pt}{1pt}
\tabincell{c}{TDA-Recovery \\ decision-making approach} & $39.757$ dB & $0.113$ s \\
\midrule
\tabincell{c}{TDA-Computation \\ decision-making approach} & $39.365$ dB & $5.6$ h \\
\specialrule{0em}{1pt}{1pt}
\bottomrule
\end{tabular}
\end{table}

Hence, we propose TDA-Recovery decision-making approach. Instead of a repetitive TDA computation or complete image recovery, we only try to establish the relationship between the transmitted topological features and source images and adopt an MLP module as the TDA decoder $\tda_{\de}$ to acquire $\hat{x}_{\tda}(\cdot)$ from the noisy $\hat{x}^{\prime}_{\tda}$, that is,
\begin{equation}
\hat{x}_{\tda} = \tda_\de(\hat{x}^{\prime}_{\tda};\phi) = \mlp(\hat{x}^{\prime}_{\tda};\phi).
\label{eq_tda_decoder}
\end{equation}

In the end, the output of TDA decoder $\tda_\de(\cdot)$ is discriminated to yield the re-transmission decision. As shown in Fig.~\ref{fig_IK_HARQ}, we design a decision network $\dec(\cdot)$, which is based on a pre-trained VGG16 network \cite{Simonyan15vgg, li2019siamvgg}, to capably measure the similarity between $\hat{x}_{\tda}$ and $\hat{x}$. Finally, the decision network $\dec(\cdot)$ in Eq.~\eqref{eq_ACK} is accomplished as
\begin{equation}
\zeta = 
\begin{cases}
1,  & \text{if} \ {\rm VGG}(\hat{x},\hat{x}_{\text{TDA}}) \geq \chi_x; \\
0, & \text{if} \ {\rm VGG}(\hat{x},\hat{x}_{\text{TDA}}) < \chi_x.   
\end{cases}   
\label{eq_ack}
\end{equation}
In essence, the re-transmission decision by $\dec(\cdot)$ is based on the assessment of whether $\hat{x}$ and $\hat{x}_{\tda}$ fall into the same category, while ${\rm VGG}(\cdot)$ outputs the semantic similarity between the received content and the TDA-induced image. This re-transmission determination can be reached if the semantic similarity between $\hat{x}$ and $\hat{x}_{\tda}$ by ${\rm VGG}(\cdot)$ exceeds the decision threshold $\chi_x$. In other words, $\hat{x}$ and $\hat{x}_{\tda}$ are sufficiently different.

Table \ref{tab_RT} compares the peak signal-to-noise ratio (PSNR) and running time between the TDA-Recovery and TDA-Computation decision-making approaches of CIFAR10 dataset under AWGN channel. The SNR is set to $10$ dB and the compression rate is set to $1/3$. The experiments are implemented in PyTorch and run on NVIDIA GeForce RTX 4090. As indicated in Table \ref{tab_RT}, while both methods yield similar PSNR performance, the former operates on a second-level time, whereas the latter requires an hour-level time. Therefore, it validates the effectiveness of the mentioned feature selection in terms of PSNR and computational costs.

\section{Experimental Setting and Numerical Results}
\label{sec:results}
\begin{table*}[!tb]
\caption{\label{HP}{The default hyper-parameter settings of swin transformer.}}
\centering
\begin{tabular}{cccccc}
\toprule
\specialrule{0em}{1pt}{1pt}  
\textbf{Hyperparameter} & No. of Stages & Depth (No. of swin transformer blocks) & Embedding dimension of each stage & No. of heads & Window size\\
\midrule
\specialrule{0em}{1pt}{1pt}
\textbf{CIFAR10} & 2 & $[2,4]$ & $[128,256]$ & $[4,8]$ & $2$ \\
\specialrule{0em}{1pt}{1pt}
\textbf{DIV2K} & 4 & $[2,2,6,2]$ & $[128,192,256,320]$ & $[4,6,8,10]$ & $8$ \\
\bottomrule
\end{tabular}
\end{table*}

\begin{table*}[!tb]
\caption{\label{HP-tda-encoder}{The default hyper-parameter settings of TDA encoder.}}
\centering
\begin{tabular}{cc}
\toprule
\specialrule{0em}{1pt}{1pt}  
\textbf{Filtration} & Values\\
\midrule
\specialrule{0em}{1pt}{1pt}
Height Filtration $\psi$ & $(1,0)$, $(0,1)$, $(1,1)$, $(1,-1)$, $(-1,1)$, $(-1,-1)$, $(-1,0)$, and $(0,-1)$ \\
\specialrule{0em}{1pt}{1pt}
Radial Filtration $(u,v)$ & $(23, 7)$, $(23, 15)$, $(23, 23)$, $(7, 23)$, $(7, 15)$, $(7, 7)$, $(15, 23)$, $(15, 15)$, and $(15, 7)$ \\
\bottomrule
\end{tabular}
\end{table*}

\subsection{Experimental Setting}
\label{sec:simulation setup}
We evaluate the performance of SC-TDA-HARQ on transmitting typical image datasets with resolutions spanning from $32 \times 32$ up to 2K. Specifically, we employ the CIFAR10 dataset for training and testing of lower resolution images. Conversely, for high-resolution images, the DIV2K dataset is selected for training purposes, while the Kodak dataset is utilized for testing. Moreover, we qualify the performance using both the widely used pixel-wise metric PSNR and the perceptual metric multi-scale structural similarity (MS-SSIM). The performance comparison is primarily conducted against varying SNRs and compression rates. In terms of setting the compression rate, it is regulated by the compression dimension $C$. For the CIFAR10 dataset, $C$ is selected from the set $\{8,16,24,32,48\}$.~Meanwhile, for high-resolution datasets, $C$ is chosen from a different set comprising of $\{16,32,64,96,128,196\}$. SC-TDA-HARQ is compared with separation-based source and channel coding (i.e., BPG codec for source image compression and LDPC codec for channel coding), Deep JSCC \cite{bourtsoulatze2019deep} and WITT \cite{yang2023witt}. For the LDPC codec, various combinations of code rates and modulations are evaluated.~In particular, $(3072, 6144)$, $(3072, 4608)$, and $(1536, 4608)$ are chosen for $1/2$, $2/3$, and $1/3$ code rates, respectively, while BPSK, 4-QAM, 16-QAM, or 64-QAM are considered for modulations \cite{yoo2023role}.

During the training, images are randomly cropped into $256 \times 256$ patches. Besides, we exploit the Adam optimizer with a learning rate of $1 \times 10^{-4}$, and the batch size is set to $128$ and $16$ for the CIFAR10 dataset and DIV2K dataset, respectively. Other essential settings for swin transformer in SC-TDA-HARQ are summarized in Table \ref{HP}.

\begin{figure}[!tb]
\centering
\includegraphics[width=.48\textwidth]{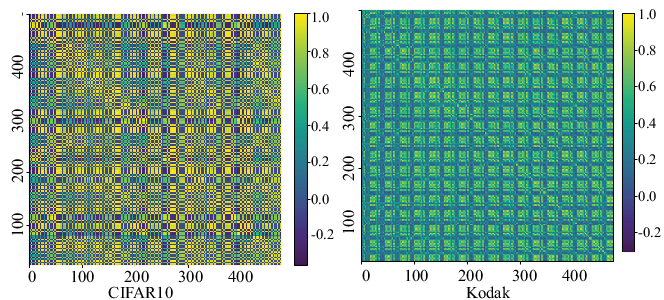}
\vspace{-.5cm}
\caption{The Pearson correlation matrix of CIFAR10 and Kodak datasets.}
\label{fig_correlation_matrix}
\end{figure}

On the other hand, the direction $\psi$ used in the height filtration and the center used in the radical filtration are set as Table \ref{HP-tda-encoder}, while typical settings for computing distance-based metrics or kernels are given in Section~\ref{TDA_HARQ}. Accordingly, the TDA encoder can generate $476$ dimensions of topological features.~In terms of Pearson coefficients, Fig.~\ref{fig_correlation_matrix} presents the correlation of these features \cite{cohen2009pearson}, $28$ of which with least correlations are selected to reduce the computational complexity without sacrificing the performance. 

\begin{figure*}[!htb]
\centering
\includegraphics[width=\textwidth]{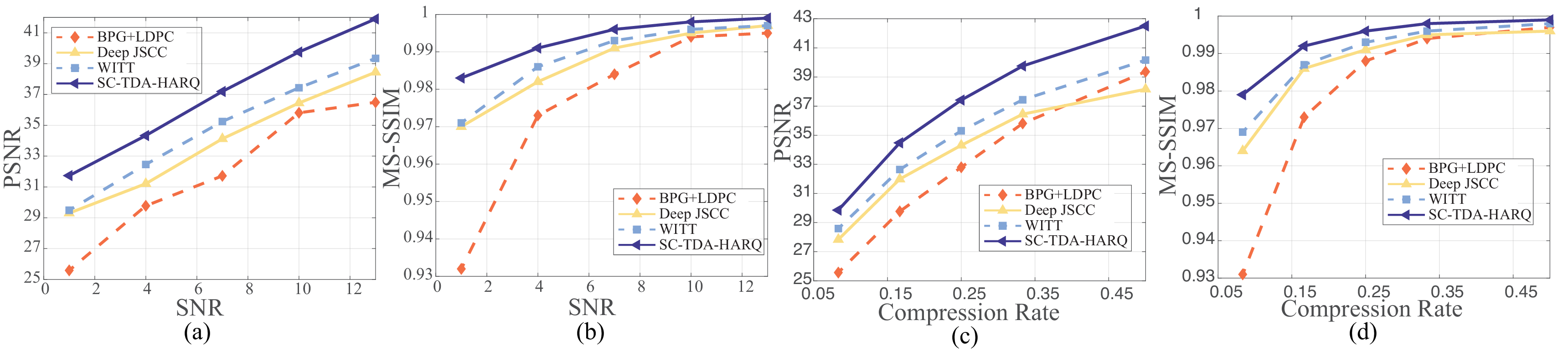}
\vspace{-0.8cm}
\caption{PSNR and MS-SSIM performance of the CIFAR10 dataset over the \textbf{AWGN} channel. (a)$-$(b) PSNR and MS-SSIM performance of the CIFAR10 dataset versus the SNR. The compression rate is set to $1/3$. (c)$-$(d) PSNR and MS-SSIM performance of the CIFAR10 dataset versus the compression rate. The SNR is set to $10$ dB.}
\label{fig_psnr-cifar10-awgn}
\end{figure*}

\begin{figure*}[!htb]
\centering
\includegraphics[width=\textwidth]{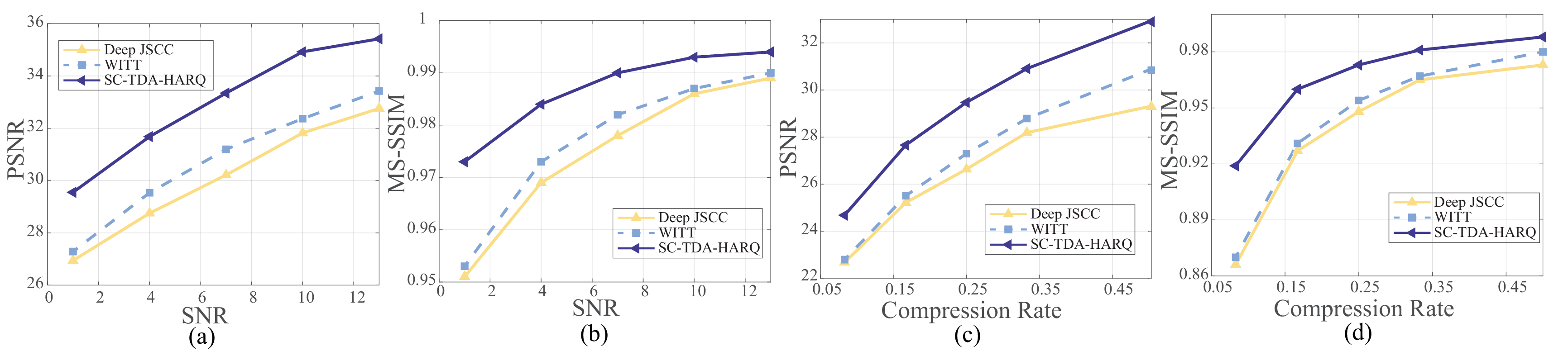}
\vspace{-0.8cm}
\caption{PSNR and MS-SSIM performance of the CIFAR10 dataset over the \textbf{Rayleigh} channel. (a)$-$(b) PSNR and MS-SSIM performance of the CIFAR10 dataset versus the SNR, respectively. The compression rate is set to $1/3$. (c)$-$(d) PSNR and MS-SSIM performance of the CIFAR10 dataset versus the compression rate. The SNR is set to $3$ dB.}
\label{fig_psnr-cifar10-rayleigh}
\end{figure*}

\begin{figure*}[!htb]
\centering
\includegraphics[width=0.8 \textwidth]{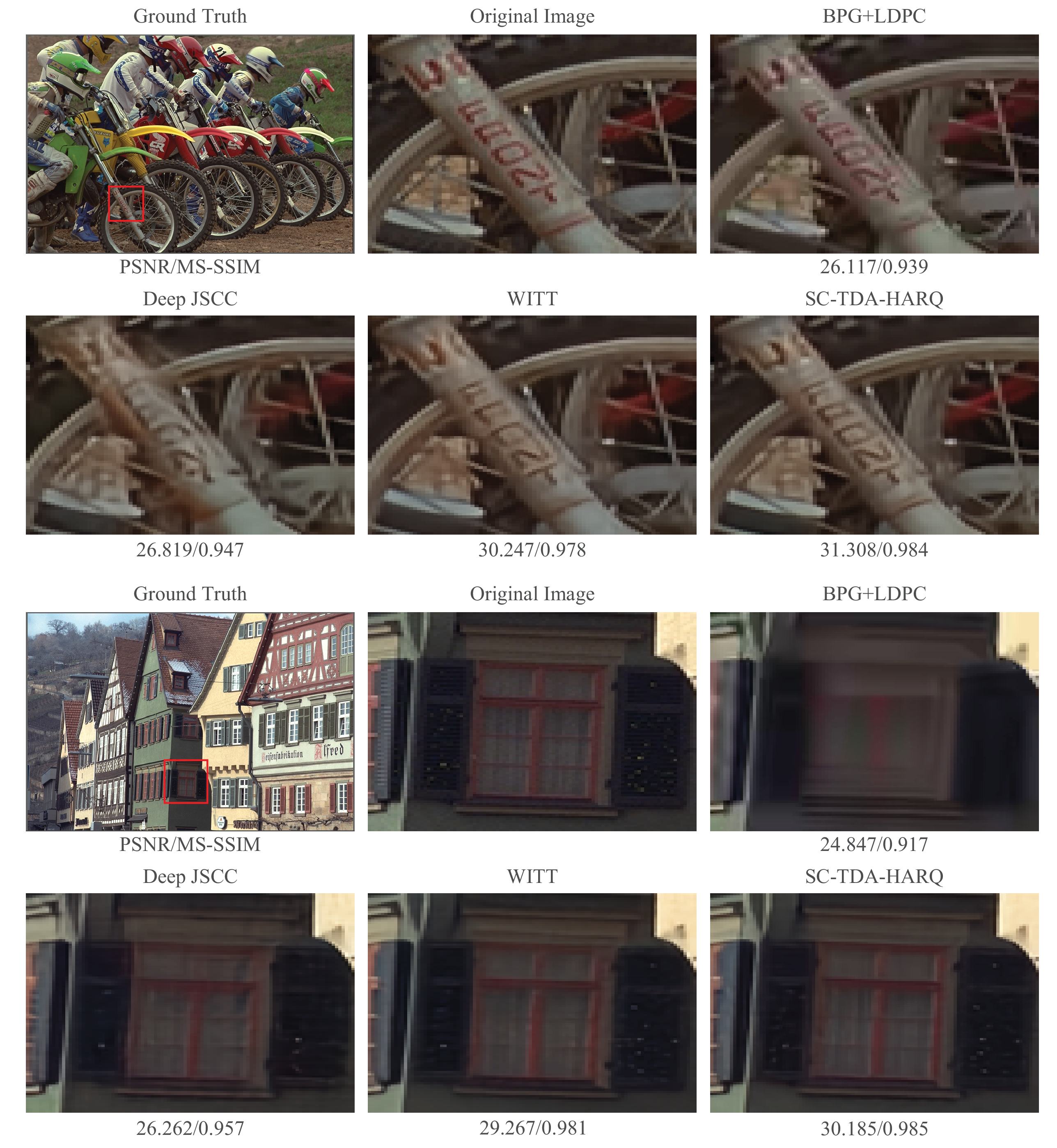}
\caption{Examples of visual comparison of Kodak dataset under AWGN channel at SNR $=10$ dB and $R=1/16$.}
\label{fig_visualization}
\end{figure*}

\begin{figure*}[!htb]
\centering
\includegraphics[width=\textwidth]{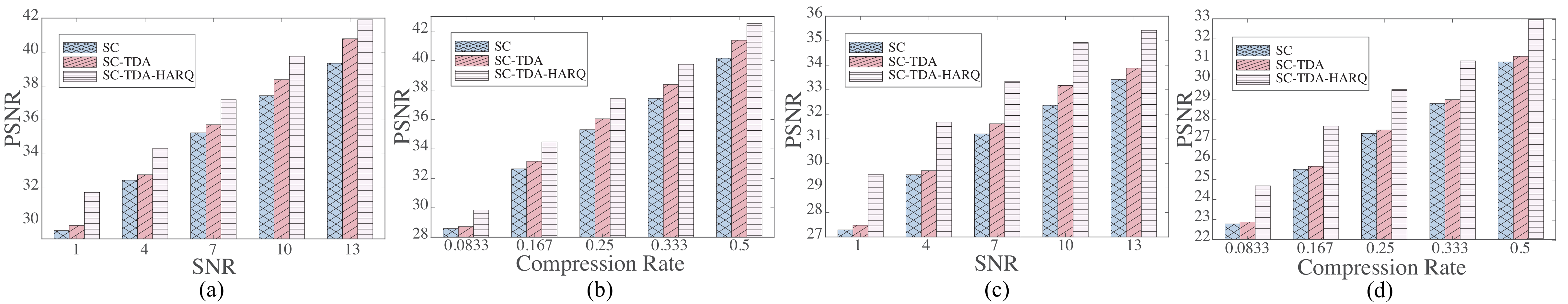}
\vspace{-0.8cm}
\caption{The PSNR comparison of the \textbf{CIFAR10} dataset versus the SNR and compression rate with the assistance of TDA. (a) The compression rate is set to $1/3$ under AWGN channel. (b) The SNR is set to $10$ dB under AWGN channel. (c) The compression rate is set to $1/3$ under Rayleigh channel. (d) The SNR is set to $3$ dB under Rayleigh channel.}
\label{fig_impact-tda}
\end{figure*}

\begin{figure*}[!htb]
\centering
\includegraphics[width=\textwidth]{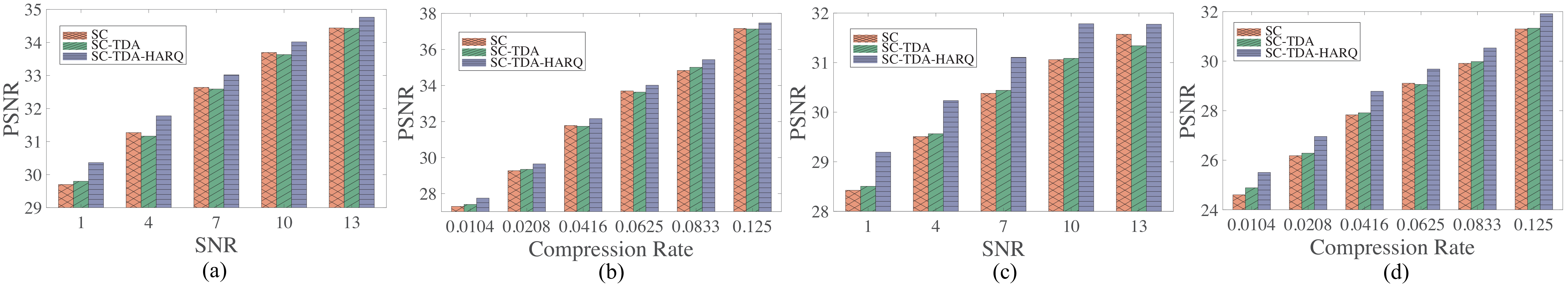}
\vspace{-0.8cm}
\caption{The PSNR comparison of the \textbf{Kodak} dataset versus the SNR and compression rate with the assistance of TDA. (a) The compression rate is set to $1/16$ under AWGN channel. (b) The SNR is set to $10$ dB under AWGN channel. (c) The compression rate is set to $1/16$ under Rayleigh channel. (d) The SNR is set to $3$ dB under Rayleigh channel.}
\label{fig_impact-tda-high}
\end{figure*}

\begin{figure*}[!ht]
\centering
\includegraphics[width=\textwidth]{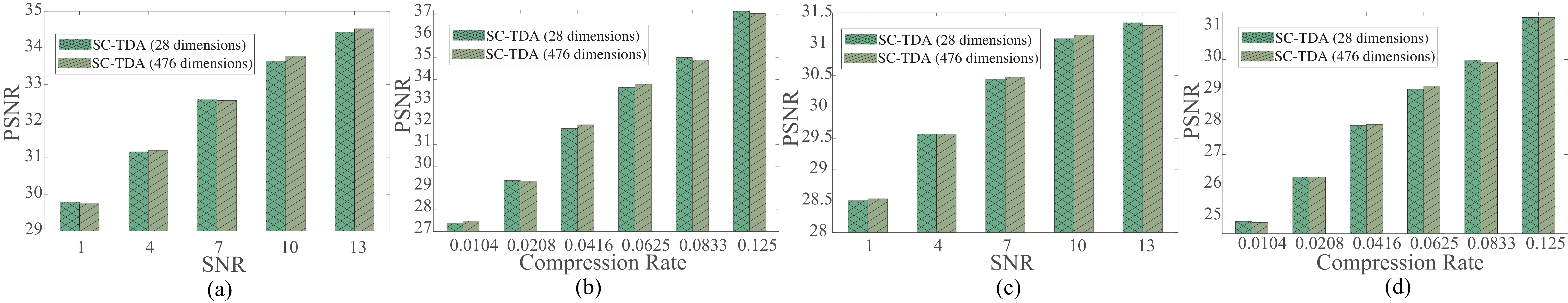}
\vspace{-0.8cm}
\caption{The PSNR comparison of the \textbf{Kodak} dataset versus the SNR and compression rate, considering different dimensions of TDA features in the \textbf{SC-TDA} model. (a) The compression rate is set to $1/16$ under AWGN channel. (b) The SNR is set to $10$ dB under AWGN channel. (c) The compression rate is set to $1/16$ under Rayleigh channel. (d) The SNR is set to $3$ dB under Rayleigh channel.}
\label{fig_tda-dimension}
\end{figure*}

\begin{figure*}[!ht]
\centering
\includegraphics[width=\textwidth]{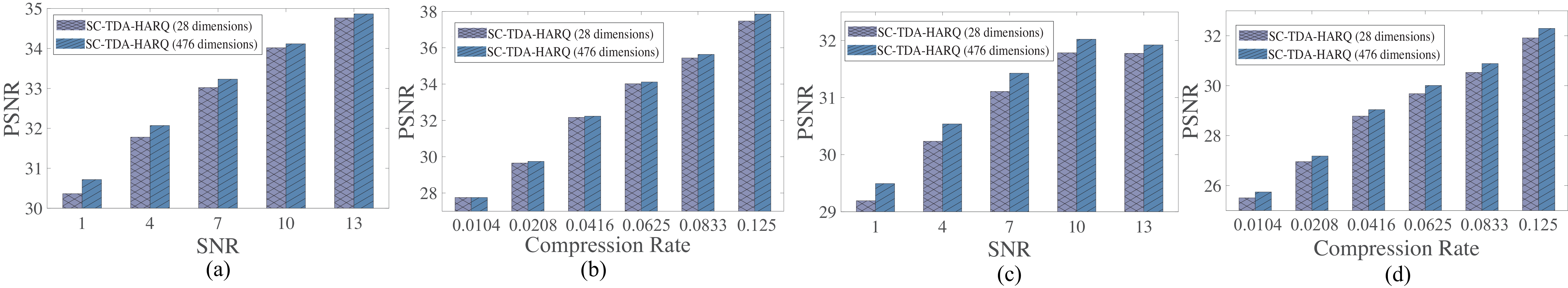}
\vspace{-0.8cm}
\caption{The PSNR comparison of the \textbf{Kodak} dataset versus the SNR and compression rate, considering different dimensions of TDA features in the \textbf{SC-TDA-HARQ} model. (a) The compression rate is set to $1/16$ under AWGN channel. (b) The SNR is set to $10$ dB under AWGN channel. (c) The compression rate is set to $1/16$ under Rayleigh channel. (d) The SNR is set to $3$ dB under Rayleigh channel.}
\label{fig_tda-dimension-harq}
\end{figure*}

\subsection{Numerical Results}
\label{sec:numerical results}
Numerically, Fig.~\ref{fig_psnr-cifar10-awgn} shows the PSNR and MS-SSIM performance of CIFAR10 dataset over the AWGN channel. Fig.~\ref{fig_psnr-cifar10-awgn}(a) and Fig.~\ref{fig_psnr-cifar10-awgn}(b) investigates the performance versus the SNR, where the compression rate is set to $1/3$.~It can be observed that for the low-resolution dataset, SC-TDA-HARQ outperforms other schemes (i.e., ``BPG+LDPC'', Deep JSCC and WITT) for all SNRs in terms of both PSNR and MS-SSIM. Meanwhile, the MS-SSIM performance gain decreases with the increase of the SNR, and SC-TDA-HARQ achieves the maximum gain when SNR is set to $1$ dB. On the other hand, Fig.~\ref{fig_psnr-cifar10-awgn}(c) and Fig.~\ref{fig_psnr-cifar10-awgn}(d) manifests the performance versus the compression rate where the SNR $=10$ dB. SC-TDA-HARQ can generally outperform both traditional and semantic communication methods by the comparison of PSNR and MS-SSIM. Moreover, the PSNR performance gap increases in direct proportion to the compression rate; while there is a significant drop in MS-SSIM performance gain with the growth of the compression rate. On the other hand, Deep JSCC achieves inferior performance than ``BPG+LDPC'' at SNR $=10$ dB, $1/2$ compression rate when transmitting CIFAR10 images.

Fig.~\ref{fig_psnr-cifar10-rayleigh} demonstrates the PSNR and MS-SSIM performance of CIFAR10 dataset over the Rayleigh channel. Fig.~\ref{fig_psnr-cifar10-rayleigh}(a) and Fig.~\ref{fig_psnr-cifar10-rayleigh}(b) investigates the performance versus the SNR, where the compression rate $R$ is set to $1/3$. By Fig.~\ref{fig_psnr-cifar10-rayleigh}(a) and Fig.~\ref{fig_psnr-cifar10-rayleigh}(b), for the low-resolution dataset, SC-TDA-HARQ is superior to conventional SemCom schemes (i.e., Deep JSCC and WITT) for all SNRs in terms of both PSNR and MS-SSIM. Different from the slight MS-SSIM gain obtained when the SNR increases to $13$ dB over the AWGN channel, SC-TDA-HARQ still holds considerable gain with the growth of the SNR over the Rayleigh channel. On the other hand, Fig.~\ref{fig_psnr-cifar10-rayleigh}(c) and Fig.~\ref{fig_psnr-cifar10-rayleigh}(d) manifests the performance versus the compression rate where the SNR $=3$ dB. Concerning the PSNR and MS-SSIM performance obtained by Deep JSCC and WITT, SC-TDA-HARQ shows substantial performance gain. There is also an enlarged gap in PSNR as the compression rate increases.~A similar observation also applies to the MS-SSIM versus the compression rate, expect that SC-TDA-HARQ yields superior significant improvement no matter how the compression rate changes.~The findings in Fig.~\ref{fig_psnr-cifar10-awgn} and Fig.~\ref{fig_psnr-cifar10-rayleigh} validate the joint advantages of TDA and IK-HARQ by making re-transmission decisions and reusing information, so as to reduce semantic misunderstanding in lossy environments. Furthermore, Fig.~\ref{fig_visualization} visualizes the reconstructions of the Kodak dataset under AWGN channel at SNR$=10$dB and the compression rate $R=1/16$. It can be observed that SC-TDA-HARQ can achieve better visual quality with the same bandwidth cost. More specifically, it avoids block artifacts and produces higher fidelity textures and details.

Next, we analyze the contributing impact of topological features in terms of PSNR. For CIFAR10 dataset, Fig.~\ref{fig_impact-tda}(a) and Fig.~\ref{fig_impact-tda}(b) demonstrates the PSNR performance versus the SNR at the compression rate $R=1/3$ and SNR $=10$ dB over the AWGN channel, respectively. When Comparing with SC, i.e., the conventional SemCom model with swin transformer-based semantic codec, it can be observed that the TDA-enhanced SemCom (SC-TDA) outperforms SC for low-resolution image datasets, which illustrates that the incorporation of TDA benefits the encoding and decoding processing during the image transmission. As the SNR increases, the performance enhancement brought about by TDA becomes more pronounced.~Meanwhile, with the fixed SNR, TDA achieves the ascending performance gap with the growth of the compression rate as well. Fig.~\ref{fig_impact-tda}(c) and Fig.~\ref{fig_impact-tda}(d) illustrate the variations of PSNR versus the SNR at the compression rate $R=1/3$ and SNR $=3$ dB over the Rayleigh channel, respectively.~Similar to the performance under AWGN channel, the incorporation of TDA obtains performance improvement to some extent, and the maximum performance gain is obtained at SNR $=10$ dB. Meanwhile, regardless of channel environments, the combination of IK-HARQ still obtains a considerable gain in the transmission of low resolution images.

On the other hand, the effects of integrating TDA features to Kodak dataset is also investigated.~Fig.~\ref{fig_impact-tda-high}(a) and Fig.~\ref{fig_impact-tda-high}(b) demonstrates the PSNR performance versus the SNR at the compression rate $R=1/16$ and SNR $=10$ dB over the AWGN channel, respectively. It can be observed that the SC-TDA mostly outperforms SC, which indicates that the incorporation of TDA benefits the encoding and decoding processing during the image transmission to some extent. Fig.~\ref{fig_impact-tda-high}(c) and Fig.~\ref{fig_impact-tda-high}(d) illustrate the PSNR performance versus the SNR at the compression rate $R=1/3$ and SNR $=3$ dB over the Rayleigh channel, respectively. Consistent with the performance under AWGN channel, the inclusion of TDA features leads to some improvement in performance, with the most significant gain observed at an SNR $=10$ dB. Moreover, Fig.~\ref{fig_impact-tda-high} shows that though SC-TDA sporadically under-performs than SC possibly limited $28$-dimensional TDA features, regardless of the channel conditions, the incorporation of IK-HARQ consistently achieves notable gains in the transmission of high-resolution images. In other words, the combination of TDA and IK-HARQ can consistently boost the transmission performance.

Next, we investigate the impact of different dimensions of TDA features on the PSNR performance. In Fig.~\ref{fig_tda-dimension}, we present a comparative analysis of PSNR performances for the Kodak dataset, using the SC-TDA model with different dimensions of TDA features.~Observations from Fig.~\ref{fig_tda-dimension} reveal that the SC-TDA model employing $28$-dimensional TDA features achieves a PSNR performance on par with that of the SC-TDA model utilizing $476$-dimensional (i.e., full-dimensional) features. Notably, this comparable performance is attained with significantly less transmission resource usage, highlighting the efficiency of the SC-TDA model when it incorporates optimally TDA features.~This finding underscores the potential of TDA features in enhancing image transmission quality while conserving transmission resources.~Additionally, in light of the instances where SC-TDA does not surpass SC, as illustrated in Fig.~\ref{fig_impact-tda-high}, Fig.~\ref{fig_tda-dimension} indicates that increasing the dimensions of the TDA features, which enhances the complexity or richness of TDA features, could address this issue to some extent. Moreover, as demonstrated in Fig.\ref{fig_tda-dimension-harq}, the SC-TDA-HARQ model consistently outperforms the SC-TDA model, and SC-TDA-HARQ with $476$-dimensional features shows slightly performance improvement over that with $28$-dimensional features.~Thus, it can be safely concluded that under sufficient length of check bits, substantial performance improvements can be anticipated.


We also compare our topological error detection method with other scheme like Sim32 in Appendix. Given the importance of establishing a suitable threshold for re-transmission by observing changes in semantic similarity, we undertake a comparative analysis. This analysis focuses on the variations in semantic similarity across different batches of images, sampled at various stages during the training process.~As is illustrated in Fig.~\ref{fig_results-sim32} and Table~\ref{table_sim32_tda}, Sim32 is less qualified for the decision of re-transmission, as the similarity computed by Sim32 remains almost unchanged under various channel conditions. On the contrary, SC-TDA-HARQ achieves more robust performance compared to Sim32, thus validating the positive effect of TDA in making the re-transmission decision. 

\section{Conclusion}
\label{sec:conclusion}
In this paper, we have proposed a high-efficiency framework SC-TDA-HARQ to improve the performance of wireless image transmission.~In particular, with the assistance of TDA, the SC-TDA-HARQ framework is built upon the swin transformer to extract hierarchical image representations. Furthermore, SC-TDA-HARQ utilizes IK-HARQ to fully use incrementally knowledge from multiple transmissions, while TDA-based error detection scheme is leveraged to determine the conditions for re-transmissions. Extensive results have demonstrated that SC-TDA-HARQ outperforms both conventional communication and classical SemCom methods, with manifested performance gain under different SNRs and compression rates. Meanwhile, due to the capability to leverage inner topological and geometric information embedded images, TDA significantly improves robustness compared to other error detection schemes like Sim32. In the future, we will investigate efficient means to compute TDA, so as to better tackle SemCom issues.

\appendix
\begin{figure*}[!tb]
\centering
\includegraphics[width=.85\textwidth]{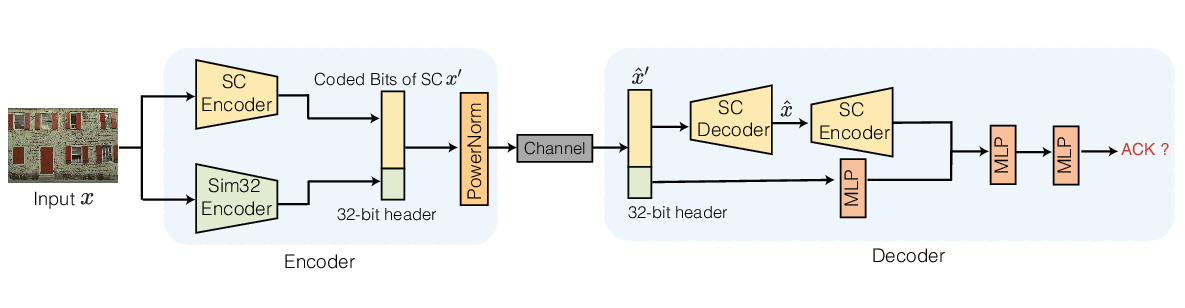}
\vspace{-0.6cm}
\caption{Pipeline of SC-HARQ with Sim32.}
\label{fig_sim32-decoder}
\end{figure*}
\textbf{Introduction of similarity computation in Fig. \ref{fig_results-sim32} and Table \ref{table_sim32_tda}}: 
As is depicted in Fig.~\ref{fig_sim32-decoder}, for Sim32, the transmission process involves two parallel encoding paths corresponding to the SC encoder and the Sim32 encoder, which are responsible for generating the coded bits for the SC and a 32-bit header, respectively \cite{jiang2022deep}. Specifically, the Sim32 encoder consists of an SC encoder-alike swin transformer and an MLP layer.~These encoded symbols are then concatenated and transmitted through the communication channel.~On the receiver side, the SC decoder reconstructs the SC coded bits. The reconstructed content is then subjected to a downsampling process using the SC encoder and combined with the symbols recovered from the $32$-bit header by the Sim32 decoder.~This amalgamated vector subsequently passes through additional MLP layers.~The final output from these MLP layers serves as an indicator of the semantic similarity between the received content and the $32$-bit header, effectively measuring how closely the received information matches the original source content \cite{jiang2022deep}.

\printbibliography
\end{document}